\documentclass[%
 reprint,
 amsmath,amssymb,
 aps,
]{revtex4-1}

\usepackage{graphicx}
\usepackage{dcolumn}
\usepackage{bm}

\begin{document}

\preprint{APS/123-QED}

\title{An Interacting Dark Energy Model with Nonminimal Derivative Coupling}

\author{Kourosh Nozari}
 \homepage{knozari@umz.ac.ir}
\author{Noushin Behrouz}
\homepage{n.behrooz@stu.umz.ac.ir}%
\affiliation{Department of Physics, Faculty of Basic Sciences,\\
University of Mazandaran,\\
P. O. Box 47416-95447, Babolsar, IRAN}

\date{\today}

\begin{abstract}
We study cosmological dynamics of an extended gravitational theory that gravity is coupled non-minimally with derivatives
of a dark energy component and there is also a phenomenological interaction between the dark energy and dark matter.
Depending on the direction of energy flow between the dark sectors, the phenomenological interaction gets two different signs. We show that this feature affects the
existence of attractor solution, the rate of growth of perturbations and stability of the solutions. By considering
an exponential potential as a self-interaction potential of the scalar field,
we obtain accelerated scaling solutions that are attractors and have the potential to alleviate the coincidence problem. While in the absence of the nonminimal
derivative coupling there is no attractor solution for phantom field when energy transfers from dark matter to dark energy, we show
an attractor solution exists if one considers an explicit nonminimal derivative coupling for phantom field in this case of energy transfer.
We treat the cosmological perturbations in this setup with details to show that with phenomenological interaction, perturbations can grow faster than the minimal case.
\begin{description}
\item[PACS numbers]
95.35.+d , 95.36.+x , 47.10.Fg
\item[Key Words]
Dynamical Systems, Dark Energy, Dark Matter,
Non-Minimal Derivative Coupling, Statefinder Diagnostic, Cosmological Perturbations.
\end{description}
\end{abstract}

\maketitle


\section{Introduction}

Observational data such as the type Ia Supernovae
redshift-distance surveys \cite{Riess,Perlmutter,Riess1,Astier,Kowalski,planck2013,planck2015}, the Baryon Acoustic Oscillations of the matter density power spectrum \cite{Percival, planck2013, planck2015} and the angular location of the first peak in the CMB power spectrum \cite{Spergel, Komatsu, planck2013, planck2015} from various origins show that the universe currently is experiencing a positive accelerating phase of expansion. To describe this expansion, one can modify the gravitational sector \cite{Sotiriou,Nojiri,Capozziello} or modify the content of the universe by introducing a dark energy component with negative pressure that violates the strong energy condition. The cosmological constant with EoS $\omega=-1$ is the simplest model of dark energy that coincides extraordinarily with observational data, but it suffers from lake of dynamics and fine-tuning problems \cite{Sahni, Carroll, Padmanabhan, Peebles}.

The dark energy scenario can be described by various scalar fields with variety of dynamical equation of state, among them we can mention quintessence field (a canonical scalar field) ~\cite{Ratra, Wetterich, Caldwell, Zlatev}, phantom field (a scalar field with negative kinetic term) \cite{Caldwell1, Caldwell2, Nojiri1, Onemli, Copeland, Saridakis, Dutta}, a combination of both these fields in a unified model called the quintom field model ~\cite{Guo, Nozari, Nozari1}, tachyon fields that emerge from string theory \cite{Padmanabhan1, Sen, Sen1, Nozari2}, k-essence fields (with a generalized kinetic energy term) \cite{Armendariz-Picon, Chiba} and Chaplygin gas component \cite{Bilic, Karami}. Furthermore, there are more complex models that describe the dark energy, in which the fields are non-minimally coupled to the background curvature. Application of these extended scenarios, dubbed “scalar-tensor” theories \cite{Uzan, Bartolo, Bertolami, Boisseau, Faraoni, Gannouji, Gupta, Nozari3}, have interesting cosmological outcomes in both inflation and the dark energy eras. As it was shown in \cite{Burgess, Barbon, Han, Lerner}, non-renormalizable operators coming out from the non-minimal coupling violate the unitarity bound of the theory during inflation era. To avoid this unitarity violation and also to find a framework that the Higgs boson would behave like a primordial Inflaton, one can consider non-minimal coupling between the derivatives of the scalar fields and curvature \cite{Amendola, Capozziello1, Capozziello2, Sushkov, Dent, Germani}. This scenario can be regarded as a subset of the most general scalar-tensor theories. In Refs. ~\cite{Granda, Granda1} coupling between the scalar field and the kinetic term has been considered as a source of dark energy, and the role of this coupling in the late-time cosmic speed up has been investigated. These theories emerge as low energy limit of some higher dimensional theories, like superstring theory ~\cite{Green} and also appear as part of the Weyl anomaly in \emph{N} = 4 conformal supergravity \cite{Hong, Nojiri1}. Furthermore, from a perturbative viewpoint, a new window has been opened on the issue of quantum gravity proposal in this framework \cite{Donoghue}. The role of this non-minimal derivative coupling during inflation has been considered in Refs. \cite{Saridakis1, Tsujikawa, Sadjadi}.

From another perspective, possible interaction between the dark energy and dark matter opens new window on the issue of the cosmological coincidence problem. Although there is no direct evidence for interaction between the dark sectors at least currently, in the absence of a fundamental theory that excludes interaction between the dark sectors, we can consider non-minimal interaction between dark energy and dark matter to alleviate coincidence problem \cite{Amendola1, Zimdahl, Amendola2, Gonzalez, Boehmer, Behrouz, Bolotin}. Moreover this interaction potentially improves interpretation of observational data \cite{planck2015de, Valiviita, Richarte}. Therefore, it is important, at least theoretically, to see possible outcomes of such an interaction and its impact on late time cosmological dynamics. For this reason we include also an interaction, much on the basis of some phenomenological considerations, between the dark sectors with the hope to shed some light on the issue of cosmological coincidence problem. By considering such an interaction between the dark sectors, whether the energy flows from dark matter to the dark energy or the reverse occurs, now is an important issue in late time cosmic dynamics. The direction of energy flow due to interaction between the dark sectors affects considerably the issues such as the existence of attractor solutions, growth rate of perturbations and the stability of cosmological solutions. With these points in mind, we consider two different candidates for dark energy: a quintessence and a phantom field, and in each case we analyze the cosmological dynamics in phase space, the statefinder diagnostic, stability in $w-w'$ phase plane and the full analysis of the perturbations in this setup with some exact solutions. The behavior of these solutions for matter perturbations on sub-Hubble scales are treated carefully for matter and scaling solutions eras.

\section{The Setup}

We consider an extension of scalar-tensor theories of gravity that
derivatives of a scalar field, as a dark energy candidate, are coupled to curvature
and there is also a phenomenological interaction between the dark energy and dark matter components. Our final goal with these types of extension is to see the status of coincidence
problem and also growth rate of perturbations in this setup.  Following the pioneer work of Amendola \cite{Amendola}, the Lagrangian of possible interaction between gravity and derivatives of the dark energy component can be sorted as follows
\begin{eqnarray}
&&L_{1}=k_{1}R\varphi_{,\mu}\varphi^{,\mu},~L_{2}=k_{2}R_{\mu\nu}\varphi^{,\mu}\varphi^{,\nu},~L_{3}=k_{3}R\varphi\Box\varphi,\nonumber\\
&&L_{4}=k_{4}R_{\mu\nu}\varphi\varphi^{;\mu\nu},~L_{5}=k_{5}R_{;\mu}\varphi\varphi^{,\mu},~~L_{6}=k_{6}\Box R\varphi^2.\nonumber
\end{eqnarray}
(for more details see also \cite{Capozziello2}). Here we just consider $L_{1}$ and $L_{2}$ since, as discussed
in \cite{Amendola, Capozziello1, Sushkov, Dent, Saridakis1}, using total divergences and without loss
of generality one can keep only the first two terms. The coefficients $k_{1}$ and $k_{2}$ are coupling parameters with dimension of
length-squared. As a specific case, and more importantly in order the resulting theory to be free of ghosts (see for instance \cite{Dent}), we set $k_{2}=-2k_{1}=\eta$, which gives the Einstein tensor $G_{\mu\nu}$. Therefore, the ghost-free action of our setup takes the following form
\begin{eqnarray}\label{action}
{\cal{S}}=\int d^{4}x\sqrt{-g}\Bigg[\frac{R}{2}
 -\frac{1}{2}(\epsilon g_{\mu\nu}-\eta G_{\mu\nu})\partial^{\mu}\varphi\partial^{\nu}\varphi\nonumber\\
 -V(\varphi)-F(\varphi){\cal{L}}_m\Bigg].
\end{eqnarray}
where $F(\varphi)=\beta e^{\alpha\varphi}$ is the interacting term between the dark sectors with constants $\alpha$ and $\beta>0$, and
$R$ is the curvature scalar, $\varphi$ is the homogeneous scalar field (as a dark energy component), $V(\varphi)$ is the scalar field
potential and ${\cal{L}}_m$ is the Lagrangian density of matter (all sorts of matter except baryons and radiation which are subdominant and supposed to be minimally coupled to gravity).
We consider the system of units in which $8\pi G = c = \hbar = 1$. In addition, we use a symbol $\epsilon$ in order to show quintessence
and phantom field in a unified manner so that $\epsilon$ takes the value $+1$ for the quintessence field and $-1$ for the phantom field.
By taking variation of the action \eqref{action} with respect to the metric, we get the field equations \cite{Sushkov, Dent} as follows

\begin{eqnarray}
G_{\mu\nu}=\epsilon T_{\mu\nu}^{(\varphi)}+\eta T_{\mu\nu}^{(\eta)}+T_{\mu\nu}^{(m)}-g_{\mu\nu}V(\varphi).
\end{eqnarray}
with

\begin{equation}\label{T1}
T_{\mu\nu}^{(\varphi)}=\nabla_{\mu}\varphi\nabla_{\nu}\varphi-\frac{1}{2}g_{\mu\nu}(\nabla{\varphi})^2,
\end{equation}

\begin{eqnarray}\label{T2}
&&-T_{\mu\nu}^{(\eta)}=-\frac{1}{2}\nabla_{\mu}\varphi\nabla_{\nu}\varphi R+2\nabla_{\alpha}\varphi\nabla_{(\mu}\varphi R^{\alpha}_{\nu)}-\nabla_{\mu}\nabla_{\nu}\varphi\Box\varphi\nonumber\\
&&+\nabla^{\alpha}\varphi\nabla^{\beta}\varphi R _{\mu\alpha\nu\beta}+\nabla_{\mu}\nabla^{\alpha}\varphi\nabla_{\nu}\nabla_{\alpha}\varphi-\frac{1}{2}(\nabla\varphi)^2G_{\mu\nu}+\nonumber\\
&&g_{\mu\nu}\Big[-\frac{1}{2}\nabla^{\alpha}\nabla^{\beta}\nabla_{\alpha}\nabla_{\beta}\varphi+\frac{1}{2}(\Box\varphi)^2
-\nabla_{\alpha}\varphi\nabla_{\beta}\varphi R^{\alpha\beta}\Big].
\end{eqnarray}
where $\nabla_{(\mu}\varphi R^{\alpha}_{\nu)}=\frac{1}{2}(\nabla_{\mu}\varphi R^{\alpha}_{\nu}+\nabla_{\nu}\varphi R^{\alpha}_{\mu})$
and $T_{\mu\nu}^{(\varphi)}$, $T_{\mu\nu}^{(\eta)}$ correspond to the variation of the terms that depend on the scalar field
in the Jordan frame and $T_{\mu\nu}^{(m)}$ is the ordinary energy-momentum tensor of matter component.
Considering a spatially-flat Friedmann-Robertson-Walker metric as

$$ds^{2}=-dt^{2}+a^{2}(t)(dr^{2}+r^{2}d\Omega^{2}),$$
\begin{equation}
d\Omega^{2}=d\theta^2+\sin ^2\theta d\varphi^2.
\end{equation}
where $t$ is the cosmic time, ($r$, $\theta$, $\varphi$) are the comoving spatial
(radial and angular) coordinates, $a(t)$ is the scale factor and $H=\dot{a}/a$ is the Hubble parameter,
 the field equations \eqref{T1} and \eqref{T2} for (00) and
(11) components (energy density and pressure, respectively) take the following form

\begin{equation}\label{rho}
\rho_\varphi=\epsilon\frac{1}{2}\dot{\varphi}^2+V(\varphi)+\frac{9}{2}\eta H^2\dot{\varphi}^2,
\end{equation}

\begin{equation}\label{p}
p_\varphi=\epsilon\frac{1}{2}\dot{\varphi}^2-V(\varphi)-\eta (\dot{H}\dot{\varphi}^2+\frac{3}{2}H^2\dot{\varphi}^2+2H\dot{\varphi}\ddot{\varphi}).
\end{equation}
Friedmann equations can be written as

\begin{equation}\label{f}
3H^2=F(\varphi)\rho_m+\frac{1}{2}\dot{\varphi}^2(\epsilon+9\eta H^2)+V(\varphi),
\end{equation}
\begin{equation}\label{ac}
\dot{H}(1-\frac{1}{2}\eta\dot{\varphi}^2)=-\epsilon\frac{1}{2}\dot{\varphi}^2-
\frac{1}{2}\gamma F(\varphi)\rho_{m}-\eta(\frac{3}{2}H^2\dot{\varphi}^2-H\dot{\varphi}\ddot{\varphi}).
\end{equation}
where $\gamma\equiv1+w_m$ is the barotropic index which depends on
the type of matter. Variation of the action \eqref{action} with respect to the scalar field gives the equation of motion
of this field as

\begin{equation}\label{motion}
\epsilon(\ddot{\varphi}+3H\dot{\varphi})+3\eta(H^2\ddot{\varphi}+2\dot{\varphi}H\dot{H}+3H^3\dot{\varphi})+V'(\varphi)=
F'(\varphi)\rho_{m}\,,
\end{equation}
where a prime represents derivative with respect to $\varphi$. The continuity equations for scalar field and dark matter are respectively as follows

\begin{equation}\label{con}
\dot{\rho}_\varphi+3H(1+\omega_\varphi)\rho_\varphi=Q\,,
\end{equation}

\begin{equation}
(F(\varphi)\rho_m)\dot{}+3H\gamma F(\varphi)\rho_m=-Q\,.
\end{equation}
where $Q=F'(\varphi)\dot{\varphi}\rho_{m}$ is a specific interaction term obtained in this model. The sign of \emph{Q} shows the direction of energy transfer

\begin{center}
$\left\{%
\begin{array}{ll}
  Q>0 , & \hbox{Energy transfers } \\
        & \hbox {from dark matter to dark energy.} \\
  Q<0 , & \hbox{Energy transfers } \\
        & \hbox {from dark energy to dark matter.}\\
\end{array}%
\right.$
\end{center}
This sign has important role in the existence of attractor solutions, growth rate of perturbations and the stability of cosmological solutions.
In comparison with the standard continuity equation, we have

\begin{equation}\label{con}
\dot{\rho}_\varphi+3H(1+\omega_{\varphi,eff})\rho_\varphi=0\,,
\end{equation}

\begin{equation}
(F(\varphi)\rho_m)\dot{}+3H\gamma_{eff}F(\varphi)\rho_m=0\,.
\end{equation}
where

\begin{equation}
\omega_{\varphi,eff}=\omega_{\varphi}-\frac{Q}{3H\rho_{\varphi}},~~~~~~~\gamma_{eff}=\gamma+\frac{Q}{3HF(\varphi)\rho_m}.
\end{equation}

Depending on the direction of energy flow from DM to DE or vice versa, the growth rate of DM density differs from the standard case without interaction. For $Q>0$

\begin{center}
$\left\{%
\begin{array}{ll}
  Dark~Energy, &\hbox{~~~~~~~~~~~~~$\omega_{\varphi,eff}<\omega_{\varphi}$,}\\
               &\hbox{DE gets red-shifted slower than }\\
               &\hbox{~~~~~~~~~~~~~$a^{-3(1+\omega_{\varphi})}$}\\
  Dark~Matter, &\hbox{~~~~~~~~~~~~~$\gamma_{eff}>\gamma$,}\\
               &\hbox{DM gets red-shifted faster than $a^{-3}$}\\
\end{array}%
\right.$
\end{center}

For $Q<0$

\begin{center}
$\left\{%
\begin{array}{ll}
  Dark~Energy, &\hbox{~~~~~~~~~~~~~$\omega_{\varphi,eff}>\omega_{\varphi}$,}\\
               &\hbox{DE gets red-shifted faster than}\\
               &\hbox{~~~~~~~~~~~~~$a^{-3(1+\omega_{\varphi})}$}\\
  Dark~Matter, &\hbox{~~~~~~~~~~~~~$\gamma_{eff}<\gamma$,}\\
               &\hbox{DM gets red-shifted slower than $a^{-3}$}\\
\end{array}%
\right.$
\end{center}

\section{The Phase Space Analysis}

Now we focus on the cosmological status of this model via a dynamical system analysis.
This technique has the capability to shed light on the existence and stability of critical points
in the cosmic history of the model, each corresponding to a cosmological phase of expansion.
We also focus mainly on the role of the non-minimal derivative coupling and the interaction between the dark sectors in this setup.
For this purpose we introduce some new dimensionless variables to translate
our equations in the language of the autonomous dynamical system. We
consider the following dimensionless quantities

\begin{eqnarray}\label{di}
x_1=\frac{\dot{\varphi}}{\sqrt{6}H},~~x_2=\frac{\sqrt{V(\varphi)}}{\sqrt{3}H},\nonumber\\
x_3=\frac{\sqrt{F(\varphi)\rho_m}}{\sqrt{3}H},~~x_4=3\sqrt{\eta}H.
\end{eqnarray}
We obtain a constraint on the parameters space of the model by
rewriting the Friedmann equation \eqref{f} in terms of the new variables
as follows

\begin{equation}\label{constraint}
1=x_1^2(\epsilon+x_{4}^2)+x_2^2+x_3^2\,.
\end{equation}
which allows us to investigate evolution of just three variables since the
forth one can be expressed in terms of the other ones. In which follows we
consider $x_1$ as our dependent variable and omit it in our forthcoming calculations. We suppose the case with positive $Q$ where the energy
flows from dark matter to dark energy. We rewrite the Friedmann equation (\ref{ac}) and the equation of motion (\ref{motion}) versus the new phase space variables.
To this end, we consider an exponential potential as $V(\varphi)=V_0e^{-\lambda \varphi}$ where
$\lambda$ and $V_0$ are positive constants. Then we find,

\begin{eqnarray}\label{Hdot}
\frac{\dot{H}}{H^2}=&&\frac{1}{1-(\frac{1}{3}-\frac{\frac{4}{9}x_4^2}{\epsilon+\frac{1}{3}x_4^2})x_1^2x_4^2}
\Bigg[-3\epsilon x_1^2-\frac{3}{2}\gamma x_3^2-x_1^2x_4^2+\nonumber\\
&&\frac{(-2\epsilon x_1+\frac{\sqrt{6}}{3}\lambda x_2^2+\frac{\sqrt{6}}{3}\alpha x_3^2-\frac{2}{3}x_1x_4^2)x_1x_4^2}{\epsilon+\frac{1}{3}x_4^2}\Bigg],
\end{eqnarray}

\begin{equation}\label{varphidot}
\frac{\ddot{\varphi}}{H^2}=\frac{-3\sqrt{6}\epsilon x_1+3\lambda x_2^2+3\alpha x_3^2-\sqrt{6}(\frac{2}{3}\frac{\dot{H}}{H^2}+1)x_1x_4^2}
{\epsilon+\frac{1}{3}x_4^2}\,,
\end{equation}

In the next step, we introduce a new time variable $N = \ln a(t)$
which is related with the cosmic time through $dN = Hdt$, and we reach the following autonomous system of equations
\begin{equation}\label{x'2}
x'_2=-\bigg(\frac{\sqrt{6}}{2}\lambda x_1+\frac{\dot{H}}{H^2}\bigg)x_2,
\end{equation}

\begin{equation}\label{x'3}
x'_3=-\bigg(\frac{3}{2}\gamma+\frac{\sqrt{6}}{2}\alpha x_1+\frac{\dot{H}}{H^2}\bigg)x_3,
\end{equation}

\begin{equation}\label{x'4}
x'_4=\frac{\dot{H}}{H^2}x_4.
\end{equation}
where a prime denotes the derivative with respect to $N$.
Now we find the critical points (fixed points) of the model to analyze the cosmological evolution and history in
this setup. For this goal, the autonomous equations (\ref{x'2} - \ref{x'4}) are set equal to zero.
To study the stability around these fixed points we have to calculate the eigenvalues in each critical points.
These eigenvalues can be derived from the following matrix equation

\begin{equation}
\left(
        \begin{array}{ccc}
         x'_2  \\
         x'_3  \\
         x'_4
       \end{array}
       \right)=M\left(
        \begin{array}{ccc}
         x_2  \\
         x_3  \\
         x_4
       \end{array}
       \right)
       \end{equation}
where M is the Jacobian matrix that is evaluated at the fixed points as follow

\begin{equation}
M=\left(
  \begin{array}{ccc}
    \frac{\partial x'_2}{\partial x_2} & \frac{\partial x'_2}{\partial x_3} & \frac{\partial x'_2}{\partial x_4} \\
                                     &                                  &             \\
    \frac{\partial x'_3}{\partial x_2} & \frac{\partial x'_3}{\partial x_3} & \frac{\partial x'_3}{\partial x_4} \\
                                     &                                  &             \\
    \frac{\partial x'_4}{\partial x_2} & \frac{\partial x'_4}{\partial x_3} & \frac{\partial x'_4}{\partial x_4} \\
  \end{array}
\right)_{(x_2, x_3, x_4)=(x_{2c}, x_{3c}, x_{4c})}
\end{equation}
The general solution for the above system in the linear approximation is

$$x_1=A_1e^{\lambda_1N}+B_1e^{\lambda_2N}+C_1e^{\lambda_3N},$$
$$x_2=A_2e^{\lambda_1N}+B_2e^{\lambda_2N}+C_2e^{\lambda_3N},$$
\begin{equation}
x_3=A_3e^{\lambda_1N}+B_3e^{\lambda_2N}+C_3e^{\lambda_3N},
\end{equation}
where $\lambda_1, \lambda_2, \lambda_3$ and \textbf{A}, \textbf{B}, \textbf{C} are
respectively eigenvalues and eigenvectors of the $M$ matrix at the critical points.
If all real eigenvalues are negative, the fixed points will be attractors (i.e. asymptotically
stable nodes), but if these real eigenvalues are positive, the fixed points will be repellers (i.e. asymptotically unstable nodes). However, if one of the eigenvalues is negative, the fixed points will be saddle points. Furthermore, if there are complex eigenvalues, depending on the sign of the real parts, they will be stable (or unstable) spirals.

To proceed further, we consider two scalar fields, quintessence and phantom field. For quintessence field $\epsilon=+1$ and we solve the above equations. For phantom field
$\epsilon=-1$. In addition, we introduce $\omega_{tot}$ as the total equation of state parameter at the critical points

\begin{equation}\label{stat}
\omega_{totc}=\frac{P_{totc}}{\rho_{totc}}=\frac{P_{\varphi}+(\gamma-1)F(\varphi)\rho_m}{\rho_{\varphi}+\rho_m}\,,
\end{equation}
where `$c$' stands for critical point. The equation of state parameter (\ref{stat}) in terms of the dimensionless parameters can be rewritten as follows

\begin{eqnarray}
\omega_{tot}=&&\epsilon x_1^2-x_2^2+(\gamma-1)x_3^2-\nonumber\\
&&\Bigg(\frac{2}{9}\frac{\dot{H}}{H^2}x_1
+\frac{1}{3}x_1+\frac{2\sqrt{6}}{27}\frac{\ddot{\varphi}}{H^2}\Bigg)x_1x_4^2.
\end{eqnarray}
To have the possibility of accelerated expansion in this setup, $\omega_{totc}$ has to be restricted as $\omega_{totc}<-\frac{1}{3}$.
Also, according to the constraint $\Omega_\varphi+\Omega_m=1$, these parameters should satisfy the following conditions

\begin{equation}\label{inequal}
0\leq \Omega_m\leq1.
\end{equation}

\begin{equation}\label{inequal}
0\leq \Omega_\varphi=[x_1^2(\epsilon+x_{4}^2)+x_2^2]\leq1.
\end{equation}

\subsection{The phase space with a quintessence field}

Solving equations (\ref{x'2} - \ref{x'4}) with $\epsilon=+1$, we reach at seven critical points $(A, B, C, D, E, F, G)$ in our system, but the critical point $G$ is not a physically acceptable point. So, we just discuss the remaining six critical points. Furthermore, we study stability of solutions around these fixed points that is related to the form of the eigenvalues in each critical point.
The results are summarized in tables \ref{tab:1} and \ref{tab:2}. Now we focus on the properties of each critical point separately. In all calculations we consider the constraint $\gamma=1$ (a pressureless matter).

\begin{figure*}
\flushleft\leftskip0em{
\includegraphics[width=.45\textwidth,origin=c,angle=0]{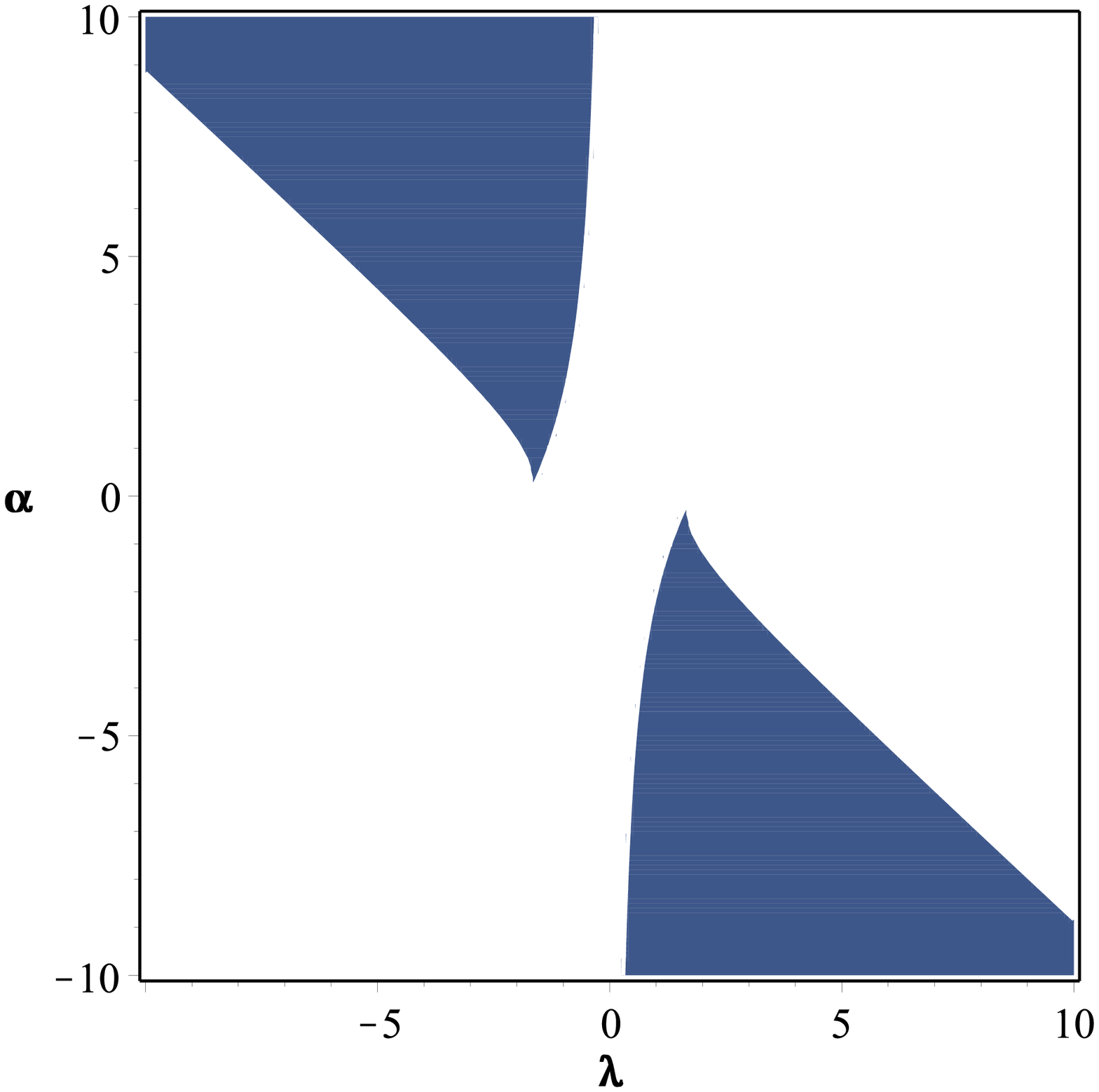}
\hspace{0.5cm}
\includegraphics[width=.45\textwidth,origin=c,angle=0]{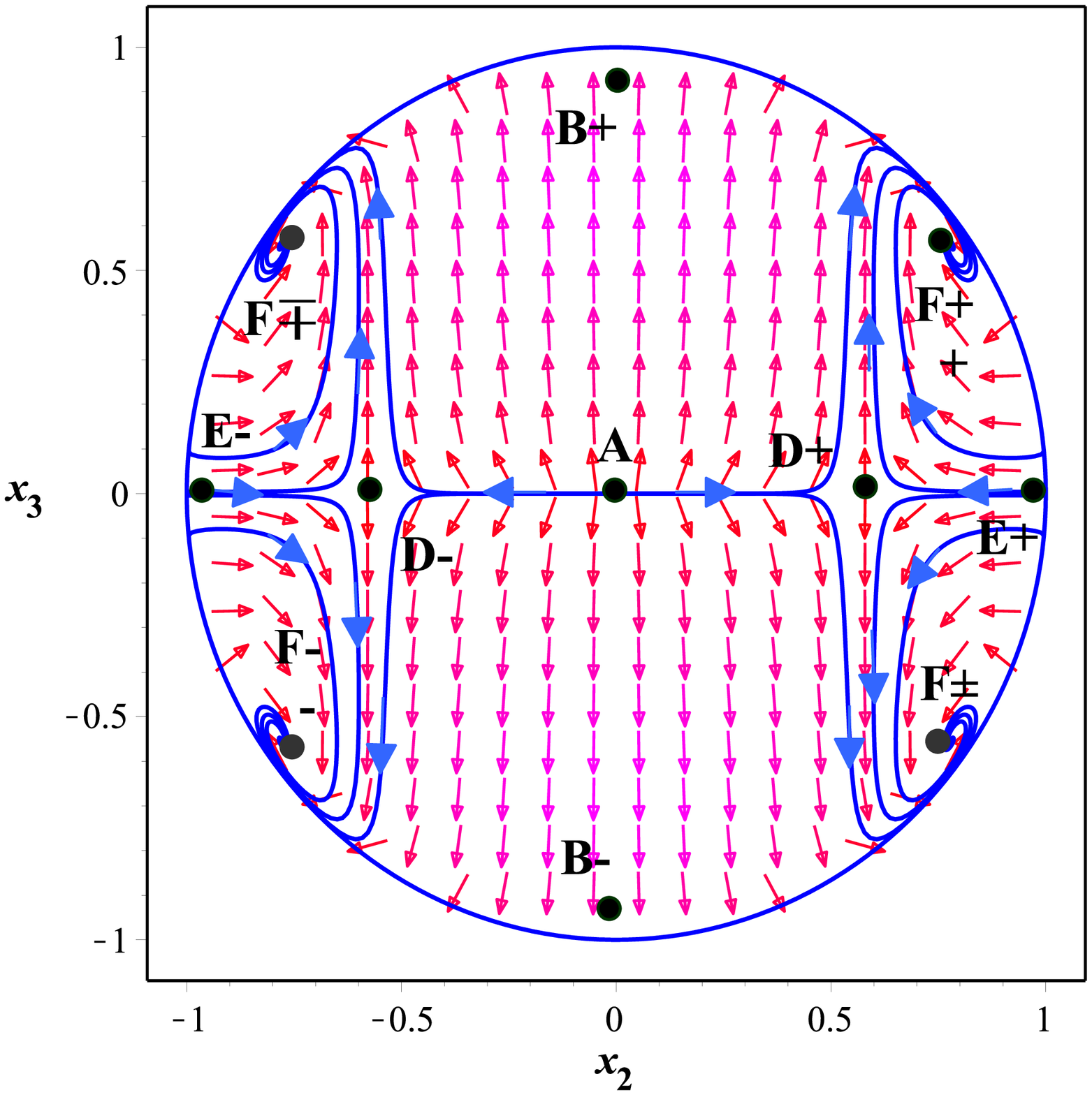}}
\caption{\label{fig1} Critical points $F_{\pm\mp}$ are stable nodes in the
blue-shaded region of the $\lambda$-$\alpha$ phase plane for quintessence field (left panel). The phase plane for
critical points $F_{\pm\mp}$ with $\lambda=2$ and $\alpha=-2.2$. The critical points $F_\pm (x_2=\pm0.78, x_3=\pm0.55)$ are stable
nodes so that the quintessence dominated solution is the
late time attractor. The critical points $A(0,~0)$, $B_{\pm}(0,\pm1)$ and $E_{\pm}(\pm1,0)$ are
saddle points. All the phase space trajectories diverge from the unstable points and
converge towards the attractors (right panel).}
\end{figure*}

\begin{itemize}
\item \textbf{Critical point A:}\\
The critical point A represents an attractor for $\alpha>\frac{\sqrt{6}}{2}$ and $\lambda>\sqrt{6}$, otherwise it is a saddle point and a scalar field's kinetic energy term dominates the universe. In this case we have no late-time acceleration.
\item \textbf{Critical points $B_\pm$:}\\
The critical points $B_\pm$ show saddle points in the phase space. These points
belong to matter domination era. The saddle nature of these points reflects the fact that the matter domination era is a transient phase in cosmic history with deceleration.
\item \textbf{Critical points $C_{\pm}$:}\\
The critical points $C_{\pm}$ show a solution with matter density and a scalar field's kinetic energy term domination. As we see, this contribution depends on the value of $\alpha$. But the behavior of these two critical points depends on the values of  $\alpha$ and $\lambda$. If we consider $\alpha^{2}<\frac{3}{2}$ and $\lambda\alpha-\alpha^2>\frac{3}{2}$, these two critical points are attractors,
otherwise they will be saddle points. Nevertheless, in both cases
there is no possibility for accelerating phase of expansion.
\item \textbf{Critical points $D_{\pm}$:}\\
The critical points $D_{\pm}$ denote either a solution with a potential
energy term domination or a scalar field's kinetic energy term domination. As
we see, this contribution depends on the values of $\lambda$. These two critical points
behave like attractor points in the phase space if we consider $\lambda^{2}<6$ and $\lambda^2-\lambda\alpha<3$.
But if we consider $\lambda^2-\lambda\alpha>3$, these two fixed points will be saddle. In both of these cases, by assuming
$\lambda^{2}<2$, accelerating phase of expansion is possible.
\item \textbf{Critical points $E_\pm$:}\\
These critical points represent cosmological constant domination phase. Unfortunately, in this case the eigenvalues are indefinite and one cannot understand the behavior of the fixed points $E_\pm$.
\item \textbf{Critical points $F_{\pm\mp}$:}\\
The critical points $F_{\pm\mp}$ are scaling solutions with accelerated expansion and naturally the coincidence problem can be alleviate in this situation.
By choosing $\lambda$ and $\alpha$ parameters from the shaded region in the left panel of figure \ref{fig1}, the critical points will be attractor nodes. For $\frac{\alpha}{\alpha-\lambda}>\frac{1}{3}$ there is an accelerating phase of expansion. In fact, our analysis verifies that $\frac{\Omega_{m}}{\Omega_{\varphi}}<1$ and $\omega_{totc}<-\frac{1}{3}$. The phase portrait for this case is illustrated in right panel of figure $1$. This figure shows that all trajectories converge to the attractor points $F_{\pm\mp}$. By choosing suitable values of quantities $\lambda$ and $\alpha$, we obtain the current value of the dark matter density parameter, $\Omega_{m}$, that is in agreement  with the recent data from Planck 2015 \cite{planck2015}, $\Omega_{m} = 0.3089\pm0.0062$
from TT, TE, EE+lowP+lensing+ext data. Furthermore, these points represent that for positive $\lambda$ we have to consider negative $\alpha$ and coupling term behaves like a  potential function.
 \end{itemize}

\begin{table*}
\begin{small}
\caption{\label{tab:1} Properties of the critical points for quintessence field.}
\begin{tabular}{cccccc}\\
\hline\hline \\
$(x_{2c},x_{3c},x_{4c})$ & Existence & Stability & $\Omega_\varphi$ & $\omega_{totc}$ & $\ddot{a}_c>0$\\\\
\hline\\ $A(0,0,0)$ & $\forall\,\lambda,~\alpha$ & attractor point if $\alpha>\frac{\sqrt{6}}{2}$ and & 1 & 1 & No \\
&  & $\lambda>\sqrt{6}$; otherwise saddle point &  &  &  \\\\
          $B_\pm(0,\pm1,0)$ & $\forall\,\lambda,~\alpha$ & saddle point & 0 & 0 & No \\\\
&  & attractor point if $\alpha^2<\frac{3}{2}$  &  &  &  \\
$C_\pm(0,\pm\sqrt{1-\frac{2\alpha^2}{3}},0)$ & $\forall\,\lambda$ and $\alpha^2\leq\frac{3}{2}$ & and $\lambda\alpha-\alpha^2>\frac{3}{2}$; & $\frac{2}{3}\alpha^2$ & $\frac{2}{3}\alpha^2$ & No \\
                       & & otherwise saddle point  &  &  &  \\\\
&  & attractor point if $\lambda^2<6$  &  &  &  \\
     $D_\pm(\pm\sqrt{1-\frac{\lambda^2}{6}},0,0)$ & $\forall\,\alpha,~\lambda^2\leq6$ & and $\lambda^2-\alpha\lambda<3$; & 1 & $-1+\frac{1}{3}\lambda^2$ &  Yes if \\
                       & & saddle point if  $\lambda^2<6$  &  &  &$\lambda^2<2$  \\
                       & & and $\lambda^2-\alpha\lambda>3$ &  &  &  \\\\
$E_{\pm}(\pm1,0,x_4)$ & $\forall\,\lambda,~\alpha$ & undefined & 1 & -1 & Yes \\\\
$F_{\pm}(\pm\frac{\sqrt{\alpha^2-\alpha\lambda+\frac{3}{2}}}{\alpha-\lambda},$ & $0\leq\alpha^2-\alpha\lambda+\frac{3}{2}\leq(\alpha-\lambda)^2$ & attractor point (fig. \ref{fig1}, &  $\frac{\alpha^2-\alpha\lambda+3}{(\alpha-\lambda)^2}$ & $-\frac{\alpha}{\alpha-\lambda}$ & Yes if \\
            $\pm\frac{\sqrt{\lambda^2-\alpha\lambda-3}}{\alpha-\lambda},0)$ & $0\leq\lambda^2-\alpha\lambda-3\leq(\alpha-\lambda)^2$ & left panel); otherwise saddle point & & & $\frac{\alpha}{\alpha-\lambda}>\frac{1}{3}$ \\\\
$G_{\pm}(0,\pm\frac{\sqrt{-2\alpha^2-6}}{\alpha},$ & not exists & - & - & - & - \\
               $ \pm\sqrt{2\alpha^2+3})$ &  &  &  &  &  \\\\
\hline \hline
\end{tabular}
\end{small}
\end{table*}

\begin{table*}
\begin{small}
\caption{\label{tab:2} The eigenvalues ($\theta_i$'s) of the critical points for quintessence field.}
\begin{tabular}{cc}\\
\hline\hline \\
point$(x_{2c},~x_{3c},~x_{4c})$ & $\theta_1$, $\theta_2$, $\theta_3$\\\\
\hline\\ $A(0,0,0)$ & $-3,~\frac{3}{2}-\frac{\sqrt{6}}{2}\alpha,~3-\frac{\sqrt{6}}{2}\lambda$ \\\\
 $B_\pm(0,\pm1,0)$ & $-\frac{3}{2}$,~$\frac{3}{2}$,~undefined \\\\
 $C_\pm(0,\pm\sqrt{1-\frac{2\alpha^2}{3}},0)$  & $-\alpha^2-\frac{3}{2},~\alpha^2-\lambda\alpha+\frac{3}{2},~\alpha^2-\frac{3}{2}$ \\\\
 $D_\pm(\pm\sqrt{1-\frac{\lambda^2}{6}},0,0)$ & $-\frac{1}{2}\lambda^2$, $\frac{1}{2}\lambda^2-\frac{1}{2}\alpha\lambda-\frac{3}{2}$, $\frac{1}{2}\lambda^2-3$ \\\\
 $E_{\pm}(\pm1,0,x_4)$ & undefined \\\\
 $F_{\pm}(\pm\frac{\sqrt{\alpha^2-\alpha\lambda+\frac{3}{2}}}{\alpha-\lambda},\pm\frac{\sqrt{\lambda^2-\alpha\lambda-3}}{\alpha-\lambda},0)$ & $\frac{3\lambda}{2(\alpha-\lambda)}$,$~\frac{-6\alpha+\frac{3}{4}\lambda\pm\frac{3}{4}\sqrt{16\alpha^2-32\alpha\lambda+25\lambda^2-72
}}{\alpha-\lambda}$\\\\
$G_{\pm}(0,\pm\frac{\sqrt{-2\alpha^2-6}}{\alpha},$  $ \pm\sqrt{2\alpha^2+3})$& - \\\\
\hline \hline
\end{tabular}
\end{small}
\end{table*}

\subsection{The phase space with a phantom field}

\begin{figure*}
\flushleft\leftskip0em{
\includegraphics[width=.48\textwidth,origin=c,angle=0]{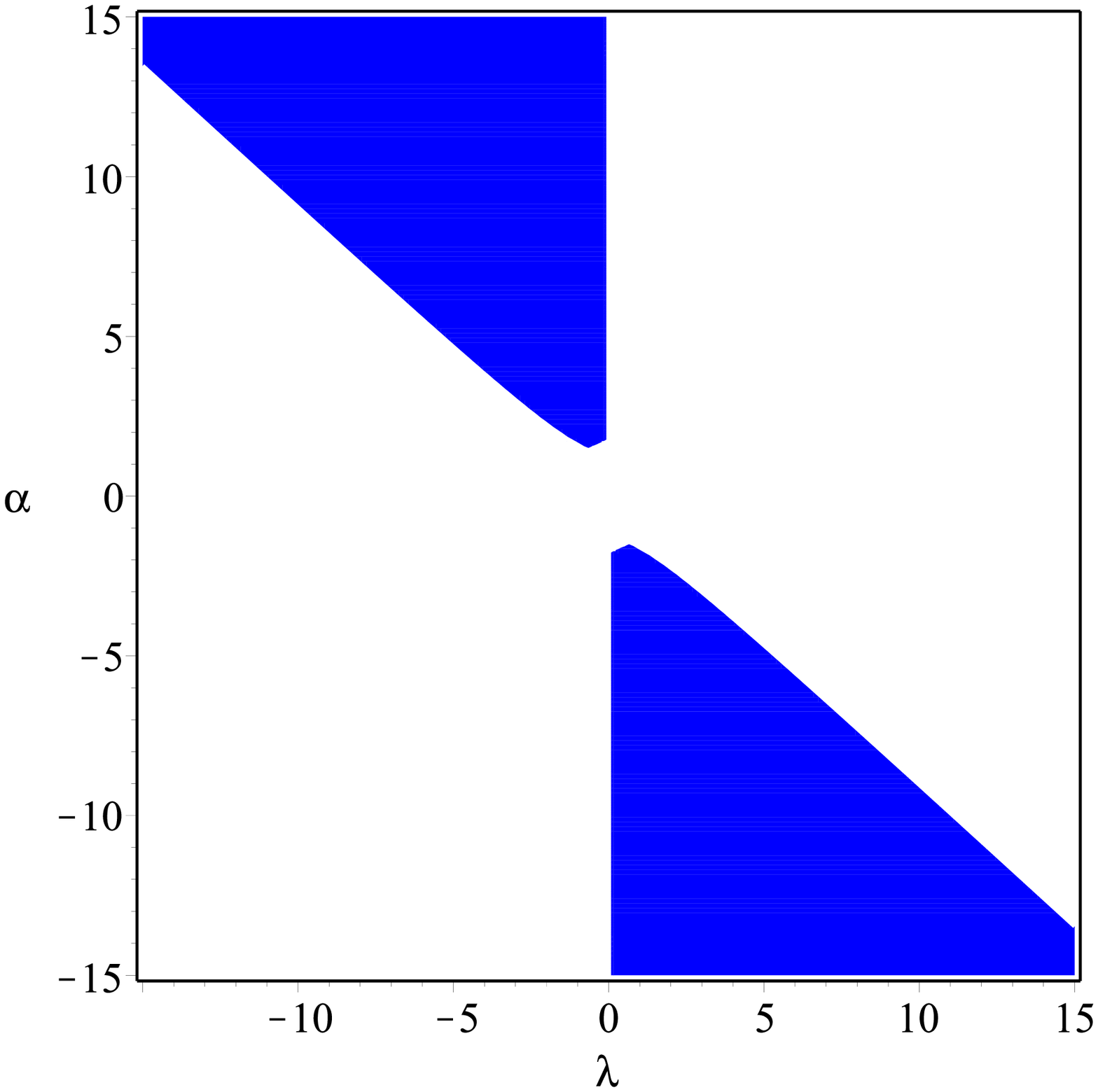}
\hspace{0.5cm}
\includegraphics[width=.48\textwidth,origin=c,angle=0]{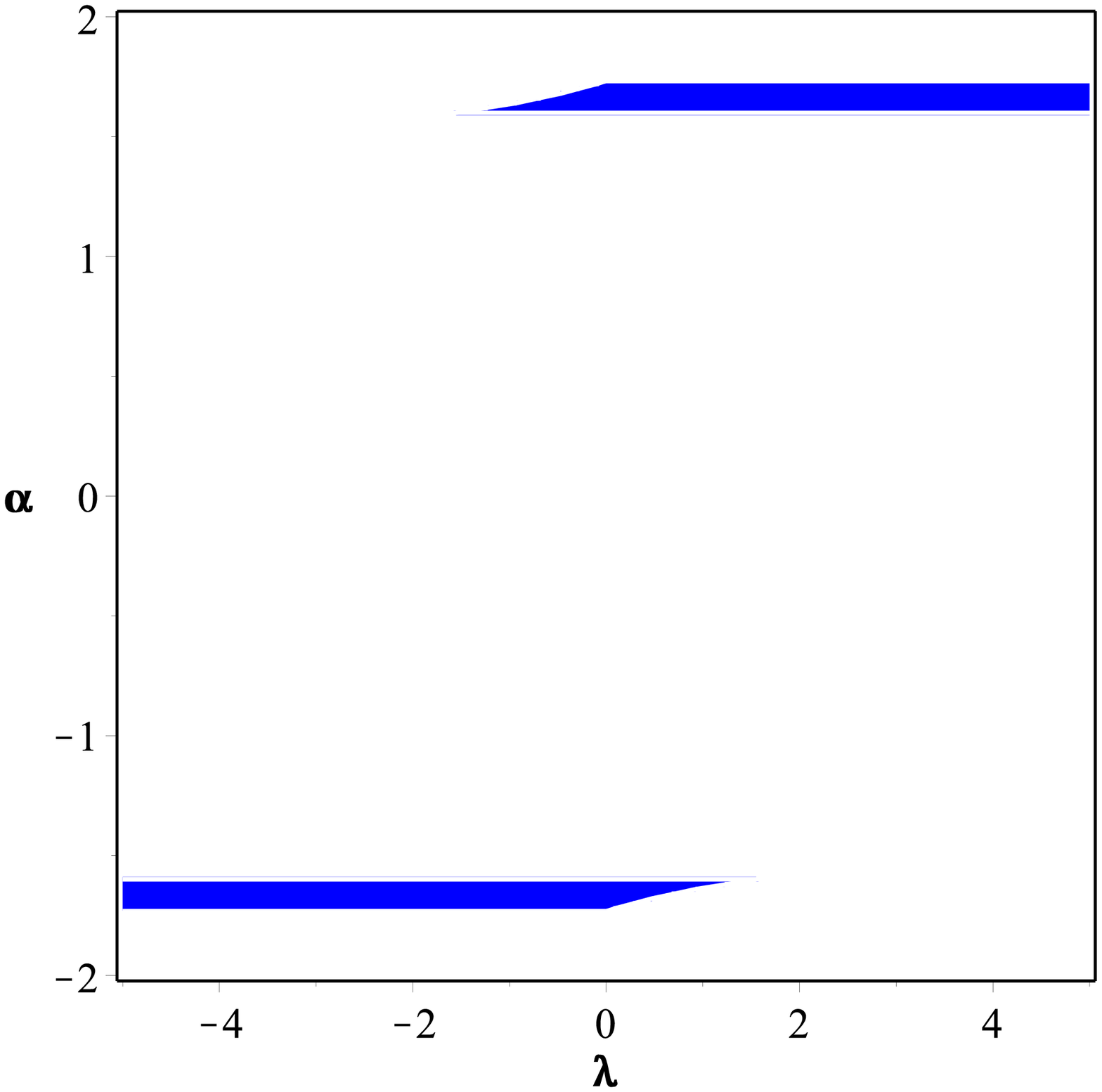}}
\caption{\label{fig2} Left panel: Critical points $F_{\pm\mp}$ are stable nodes in the
shaded region of the $\lambda$-$\alpha$ phase plane for the phantom field. Right panel: Critical points $G_{\pm\mp}$ are stable nodes in the
shaded region of the $\lambda$-$\alpha$ phase plane for the phantom field.}
\end{figure*}

Solving equations (\ref{x'2} - \ref{x'4}) with $\epsilon=-1$, we reach seven critical points $(A, B, C, D, E, F, G)$ in our system.  The critical point $C$ is not a physically acceptable point and it will not be considered in our forthcoming arguments. These critical points and stability around them are summarized in tables \ref{tab:3} and \ref{tab:4}. Now we investigate the properties of each critical point separately. As before, we suppose $\gamma=1$.
\begin{itemize}
\item \textbf{Critical point A:}\\
The fixed point $A$ shows a saddle point in a scalar field's kinetic energy dominated universe and
in this case we have no late-time acceleration.
\item \textbf{Critical points $B_\pm$:}\\
Like the previous subsection, the critical points $B_\pm$ show saddle points in the matter dominated phase and represent that
the matter domination era is transient phase.
\item \textbf{Critical points $E_\pm$:}\\
These critical points represent cosmological constant domination era. Once again, in this case the eigenvalues are indefinite and one cannot understand the behavior of the fixed points $E_\pm$.
\item \textbf{Critical points $F_{\pm\mp}$:}\\
The critical points $F_{\pm\mp}$ show either a solution with matter density term and a scalar field's kinetic energy term domination or potential term domination. According to their eigenvalues, if we choose $\alpha$ and $\lambda$ from the shaded area in the left panel of figure \ref{fig2}, the critical points will be attractor nodes, otherwise we have saddle points. There is an accelerated expansion phase under the condition $\frac{\alpha}{\alpha-\lambda}>\frac{1}{3}$. We can obtain the current value of the dark matter density, $\Omega_{m}$, that is in agreement with the recent data from Planck2015 \cite{planck2015}, $\Omega_{m} = 0.3089\pm0.0062$ by choosing $\alpha=-3.5$ and $\lambda=0.5$. Furthermore, the equation of state parameter of the dark energy gets a value very close to the equation of state parameter of the dark energy from TT, TE, EE+lowP+lensing+ext data in Ref. \cite{planck2015}, that is, $\omega=-1.019_{-0.080}^{+0.075}$. In the absence of the non-minimal derivative coupling, there is no such a good agreement with data in this setup.
The important issue about this point is that the existence of the non-minimal derivative coupling provides the possibility of having attractor solution (or scaling solution) for the present universe and this is in contrast with the previous work such as \cite{Copeland} that shows jut the future attractors. Also if we consider non-minimally coupled derivative without interaction between dark sectors, there will be no attractor points \cite{Huang}. So, we can conclude that the existence of the non-minimally coupled derivative and also interaction between dark sectors is necessary to find attractor solution.

\item \textbf{Critical points $G_{\pm\mp}$:}\\
The critical points $G_{\pm\mp}$, for a narrow range of $\alpha$, show a solution with matter density and a scalar field's kinetic energy term domination. These points also, are critical points that carry some information about the role of the non-minimal derivative coupling in this setup. These solutions are attractor if we choose $\alpha$ and $\lambda$ from the shaded region in the right panel of figure \ref{fig2}. As we see from figure $3$, $\alpha$ and $\lambda$ parameters have the same signs in contrast with the critical points $F_{\pm\mp}$. To find a value of the dark matter density parameter, $\Omega_{m}$, that is in agreement with the recent data from Planck2015 \cite{planck2015} that gives $\Omega_{m} = 0.3089\pm0.0062$, we have to consider $\alpha=1.61$ with any positive $\lambda$. Furthermore, the equation of state parameter of the dark energy reaches $-1.85$, which is close to the best fit $\omega=-1.94$ for Planck+WMAP, the best fit $\omega=-1.94$ for Planck+WMAP+high L \cite{planck2013}, $\omega=-1.54^{+0.62}_{-0.50}$ for TT and $\omega=-1.55^{+0.58}_{-0.48}$ for TE+EE in \cite{planck2015}. Our analysis shows that in the absence of the non-minimal derivative coupling, there is no such a good agreement with data in this setup. In fact, existence of a non-minimal coupling between the derivatives of the dark energy component and curvature provides a better fit with observations in this setup.
\end{itemize}

\begin{table*}
\begin{small}
\caption{\label{tab:3} Properties of the critical points for phantom field. }
\begin{tabular}{cccccc}\\
\hline\hline \\
$(x_{2c},x_{3c},x_{4c})$ & Existence & Stability & $\Omega_\varphi$ & $\omega_{totc}$ & $\ddot{a}_c>0$\\\\
\hline\\ $A(0,0,0)$ & $\forall\,\lambda,~\alpha$ & saddle point & 1 & 1 & No \\\\
 $B_\pm(0,\pm1,0)$ & $\forall\,\lambda,~\alpha$ & saddle point & 0 & 0 & No \\\\
 $C_\pm(0,\pm\sqrt{1+\frac{2\alpha^2}{3}},0)$& - & - & - & - & - \\\\
 $D_\pm(\pm\sqrt{1+\frac{\lambda^2}{6}},0,0)$ & $\forall\,\lambda,~\alpha$ & saddle point & 1 & $-1-\frac{1}{3}\lambda^2$ & Yes \\\\
 $E_{\pm}(\pm1,0,x_4)$ & $\forall\,\lambda,~\alpha$ & undefined & 1 & -1 & Yes \\\\
$F_{\pm}(\pm\frac{\sqrt{\alpha^2-\alpha\lambda-\frac{3}{2}}}{\alpha-\lambda},$ & $\alpha^2-\alpha\lambda-\frac{3}{2}\geq0$ & attractor point (fig. \ref{fig2},& $\frac{\alpha^2-\alpha\lambda-3}{(\alpha-\lambda)^2}$ & $-\frac{\alpha}{\alpha-\lambda}$ & Yes if \\
         $\pm\frac{\sqrt{\lambda^2-\alpha\lambda+3}}{\alpha-\lambda},0)$ & $0\leq\lambda^2-\alpha\lambda+3\leq(\alpha-\lambda)^2$ & left panel); otherwise saddle point & & & $\frac{\alpha}{\alpha-\lambda}>\frac{1}{3}$ \\\\
$G_{\pm}(0,\pm\frac{\sqrt{-2\alpha^2+6}}{\alpha},$ & $2\leq\alpha^2\leq3$ & attractor point (fig. \ref{fig2} right panel); & $3-\frac{6}{\alpha^2}$ & $\frac{8\alpha^4-51\alpha^2+63}{8\alpha^4-21\alpha^2+9}$ & Yes\\
       $\pm\sqrt{2\alpha^2-3})$ & & otherwise saddle point & & & \\\\
\hline \hline
\end{tabular}
\end{small}
\end{table*}

\begin{table*}
\begin{small}
\caption{\label{tab:4} The eigenvalues ($\theta_i$'s) of the critical points for phantom field($g_{i}$'s are given in Appendix A).}
\begin{tabular}{cc}\\
\hline\hline \\
point$(x_{2c},~x_{3c},~x_{4c})$ & $\theta_1$, $\theta_2$, $\theta_3$\\\\
\hline\\
$A(0,0,0)$ & $-3,~\frac{3}{2}+\frac{\sqrt{-6}}{2}\alpha,~3-\frac{\sqrt{-6}}{2}\lambda$ \\\\
 $B_\pm(0,\pm1,0)$ & $-\frac{3}{2}$,~$\frac{3}{2}$,~undefined \\\\
 $C_\pm(0,\pm\sqrt{1+\frac{2\alpha^2}{3}},0)$  & - \\\\
 $D_\pm(\pm\sqrt{1+\frac{\lambda^2}{6}},0,0)$ & $\frac{1}{2}\lambda^2$, $-\frac{5}{2}\lambda^2-9$, $-\frac{1}{2}\lambda^2-\frac{1}{2}\alpha\lambda-\frac{3}{2}$ \\\\
 $E_{\pm}(\pm1,0,x_4)$ & undefined \\\\
$F_{\pm}(\pm\frac{\sqrt{\alpha^2-\alpha\lambda-\frac{3}{2}}}{\alpha-\lambda},\pm\frac{\sqrt{\lambda^2-\alpha\lambda+3}}{\alpha-\lambda},0)$ & $\frac{3}{2}\frac{\lambda}{\alpha-\lambda}$,$~\frac{-6\alpha+\frac{3}{4}\lambda\pm\frac{3}{4}\sqrt{16\alpha^2-32\alpha\lambda
+25\lambda^2+72}}{\alpha-\lambda}$\\\\
$G_{\pm}(0,\pm\frac{\sqrt{-2\alpha^2+6}}{\alpha}, \pm\sqrt{2\alpha^2-3})$ & $g_1$, $g_2$, $g_3$ \\\\
\hline \hline
\end{tabular}
\end{small}
\end{table*}

\section{Stability in $\omega_{\varphi}'-\omega_{\varphi}$ Space}
Now we investigate the classical stability of the solutions in $\omega_{\varphi}'-\omega_{\varphi}$ phase-plane
of the scalar fields with non-minimal derivative coupling (other similar interesting cases can be seen in Refs. \cite{Caldwel3, Scherrer, Chiba1, Zhao, Nozari4, Nozari5}). Like the previous parts, a prime denotes the derivative with respect to $N = \ln a(t)$,

\begin{equation}
\omega_{\varphi}^{\prime}=\frac{d\omega_{\varphi}}{dN}=\frac{d\omega_{\varphi}}{d\rho_{\varphi}}\frac{d\rho_{\varphi}}{dN}
\end{equation}
where

\begin{equation}
\frac{d\omega_{\varphi}}{d\rho_{\varphi}}=\frac{1}{\rho_{\varphi}}\bigg(\frac{dp_{\varphi}}{d\rho_{\varphi}}-\omega_{\varphi}\bigg)
\end{equation}
and

\begin{equation}
\frac{d\rho_{\varphi}}{dN}=\frac{\dot{\rho}_{\varphi}}{H}=-3\rho_{\varphi}(1+\omega_{\varphi})
\end{equation}
By considering the sound speed as $c_{a}^2\equiv\frac{\dot{p}}{\dot{\rho}}$ or equivalently $c_{a}^2\equiv\frac{dp}{d\rho}$, we obtain the following general result
\begin{equation}
\omega'=-3(1+\omega_{\varphi})(c_{a}^{2}-\omega_{\varphi}),
\end{equation}
The sound speed represents the phase velocity of the inhomogeneous perturbations of the
scalar field. This function would be the adiabatic sound speed in this fluid if we consider the the energy-momentum of the scalar field as a perfect fluid form.  To get ride off the future big rip singularity, sound speed is supposed to be positive.
We calculate $c_{a}^{2}$ in our model and then $\omega^{\prime}$ will be achieved easily. By using equations (\ref{rho}), (\ref{p}), (\ref{con}) and the sound speed definition we get

\begin{eqnarray}
&&c_{a}^{2}=-\frac{1}{3}+
\frac{2\sqrt{6}\lambda x_{1}x_{2}^2+\frac{4\sqrt{6}}{3}\epsilon\frac{\ddot{\varphi}}{H^2}x_{1}}
{3\sqrt{6}\alpha x_1x_3^2-9\Omega_{\varphi}(1+\omega_{\varphi})}-\nonumber\\
&&\frac{\Big(\frac{2}{3}\frac{\ddot{H}}{H^2}x_{1}^{2}+
\frac{4}{9}\frac{\dot{H}}{H^2}\frac{\ddot{\varphi}}{H^2}x_{1}+\frac{2}{9}(\frac{\ddot{\varphi}}{H^2})^{2}
+\frac{2\sqrt{6}}{9}\frac{\ddot{\varphi}\dot{}}{H^3}x_{1}\Big)x_{4}^2}{3\sqrt{6}\alpha x_1x_3^2-9\Omega_{\varphi}(1+\omega_{\varphi})}
\end{eqnarray}
then we obtain the following form for $\omega'$
\begin{eqnarray}
&&\omega'=(1+\omega_{\varphi})(1+3\omega_{\varphi})-3(1+\omega_{\varphi})\nonumber\\
&&\Bigg(-\frac{\Big(\frac{2}{3}\frac{\ddot{H}}{H^2}x_{1}^{2}+
\frac{4}{9}\frac{\dot{H}}{H^2}\frac{\ddot{\varphi}}{H^2}x_{1}+\frac{2}{9}(\frac{\ddot{\varphi}}{H^2})^{2}
+\frac{2\sqrt{6}}{9}\frac{\ddot{\varphi}\dot{}}{H^3}x_{1}\Big)x_{4}^2}{3\sqrt{6}\alpha x_1x_3^2-9\Omega_{\varphi}(1+\omega_{\varphi})}\nonumber\\
&&+\frac{\frac{4\sqrt{6}}{3}\epsilon\frac{\ddot{\varphi}}{H^2}x_{1}+2\sqrt{6}\lambda x_{1}x_{2}^2-}
{3\sqrt{6}\alpha x_1x_3^2-9\Omega_{\varphi}(1+\omega_{\varphi})}\Bigg).
\end{eqnarray}
For the sake of economy we avoid to present the extended form of this equation but we note that in this relation

\begin{equation}\label{Hdot1}
\frac{\dot{H}}{H^2}=-\frac{3}{2}\Omega_{\varphi}(1+w_{\varphi}),
\end{equation}

\begin{eqnarray}
x_{2}^2=\frac{\frac{1}{2}\Omega_{\varphi}(1-\omega_{\varphi})-\frac{2}{3}x_{1}^2x_{4}^2-\frac{1}{9}\frac{\dot{H}}{H^2}x_{1}^2x_{4}^2
}{1+\frac{\frac{\sqrt{6}}{9}\lambda x_{1}x_{4}^2}{\epsilon+\frac{1}{3}x_{4}^2}}+\nonumber\\
\frac{\frac{\sqrt{6}}{27}\bigg(\frac{3\sqrt{6}\epsilon x_{1}-3\alpha x_{3}^2+\frac{2}{3}\sqrt{6}\frac{\dot{H}}{H^2}x_{1}x_{4}^2+\sqrt{6}x_{1}x_{4}^2}
{\epsilon+\frac{1}{3}x_{4}^{2}}\bigg)x_{1}x_{4}^2}{1+\frac{\frac{\sqrt{6}}{9}\lambda x_{1}x_{4}^2}{\epsilon+\frac{1}{3}x_{4}^2}}
\end{eqnarray}

By taking time derivative of equations (\ref{Hdot}) and (\ref{varphidot}) and doing some calculations, we reach the following equations respectively

\begin{eqnarray}
&&\frac{\ddot{H}}{H^3}=\frac{\frac{9}{2}\Omega_{\varphi}(1+\omega_{\varphi})-\frac{3}{2}\sqrt{6}\alpha x_1x_3^2-\frac{\sqrt{6}}{2}\epsilon \frac{\ddot{\varphi}}{H^2}x_{1}-\frac{3\sqrt{6}\lambda}{2}x_{1}x_{2}^2}
{1+(\frac{4}{9\epsilon+3x_4^2}-\frac{1}{3})x_{1}^2x_{4}^2}\nonumber\\
&&+\frac{\Big(\frac{2\sqrt{6}}{9}
\frac{\dot{H}}{H^2}\frac{\ddot{\varphi}}{H^2}x_{1}+\frac{\dot{H}}{H^2}x_{1}^2+\frac{\sqrt{6}}{6}\frac{\ddot{\varphi}}{H^2}x_{1}
+\frac{1}{9}(\frac{\ddot{\varphi}}{H^2})^2\Big)x_{4}^2}{1+(\frac{4}{9\epsilon+3x_4^2}-\frac{1}{3})x_{1}^2x_{4}^2}+\nonumber\\
&&\frac{\bigg(
\frac{2\sqrt{6}}{3}\Big(\frac{\dot{H}}{H^2}\Big)^2x_{1}+\frac{4}{3}\frac{\dot{H}}{H^2}\frac{\ddot{\varphi}}{H^2}+
3\sqrt{6}\frac{\dot{H}}{H^2}x_{1}\bigg)x_{4}^2+3\sqrt{6}\lambda^2x_{1}x_{2}^2}{\frac{9}{\sqrt{6}}\Big(\frac{1}{x_1^2x_4^2}+(\frac{4}{9\epsilon+3x_4^2}
-\frac{1}{3})x_{1}\Big)(\epsilon+\frac{1}{3}x_{4}^{2})}\nonumber\\
&&+\frac{3\sqrt{6}\epsilon \frac{\dot{H}}{H^2}x_{1}+\frac{\ddot{\varphi}}{H^2}(3\epsilon+x_{4}^2)+\alpha(3\sqrt{6}\alpha x_1x_3^2+9\gamma x_3^2)}{\frac{9}{\sqrt{6}}\Big(\frac{1}{x_1^2x_4^2}+(\frac{4}{9\epsilon+3x_4^2}
-\frac{1}{3})x_{1}\Big)(\epsilon+\frac{1}{3}x_{4}^{2})}
\end{eqnarray}

\begin{eqnarray}
&&\frac{\ddot{\varphi}\dot{}}{H^3}=\frac{-\alpha(3\sqrt{6}\alpha x_1x_3^2+9\gamma x_3^2)}{\epsilon+\frac{1}{3}x_4^2}-\nonumber\\
&&\frac{\bigg(\frac{2\sqrt{6}}{3}\Big(\frac{\dot{H}}{H^2}\Big)^2x_{1}+\frac{4}{3}\frac{\dot{H}}{H^2}\frac{\ddot{\varphi}}{H^2}+
\frac{2\sqrt{6}}{3}\frac{\ddot{H}}{H^3}x_{1}+3\sqrt{6}\frac{\dot{H}}{H^2}x_{1}\bigg)x_{4}^2}{\epsilon+\frac{1}{3}x_{4}^{2}}\nonumber\\
&&-\frac{3\sqrt{6}\lambda^2x_{1}x_{2}^2-3\sqrt{6}\epsilon \frac{\dot{H}}{H^2}x_{1}-\frac{\ddot{\varphi}}{H^2}(3\epsilon+x_{4}^2)}{\epsilon+\frac{1}{3}x_4^2}.
\end{eqnarray}

\begin{figure*}
\flushleft\leftskip0em{
\includegraphics[width=.48\textwidth,origin=c,angle=0]{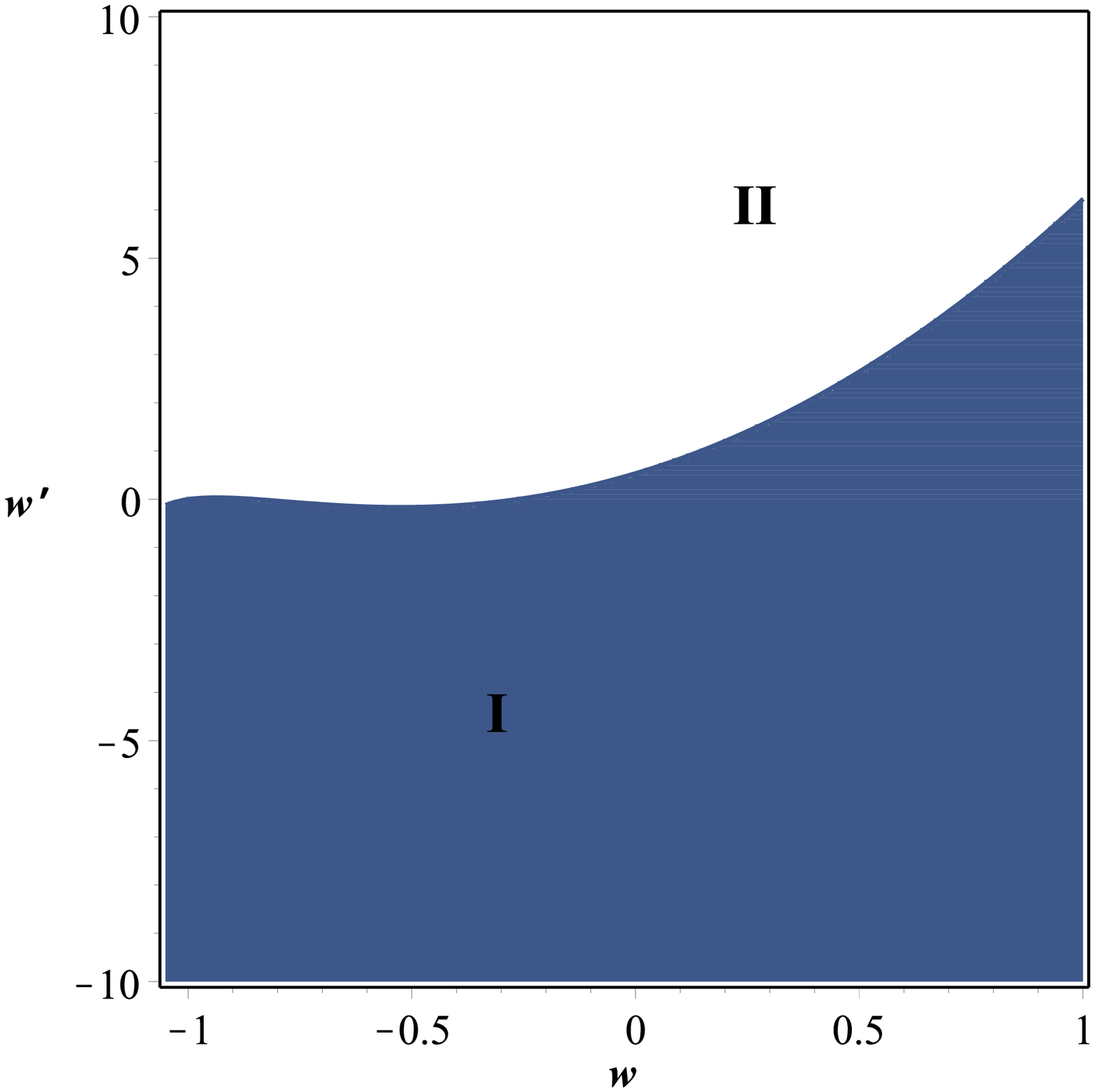}
\hspace{0.5cm}
\includegraphics[width=.48\textwidth,origin=c,angle=0]{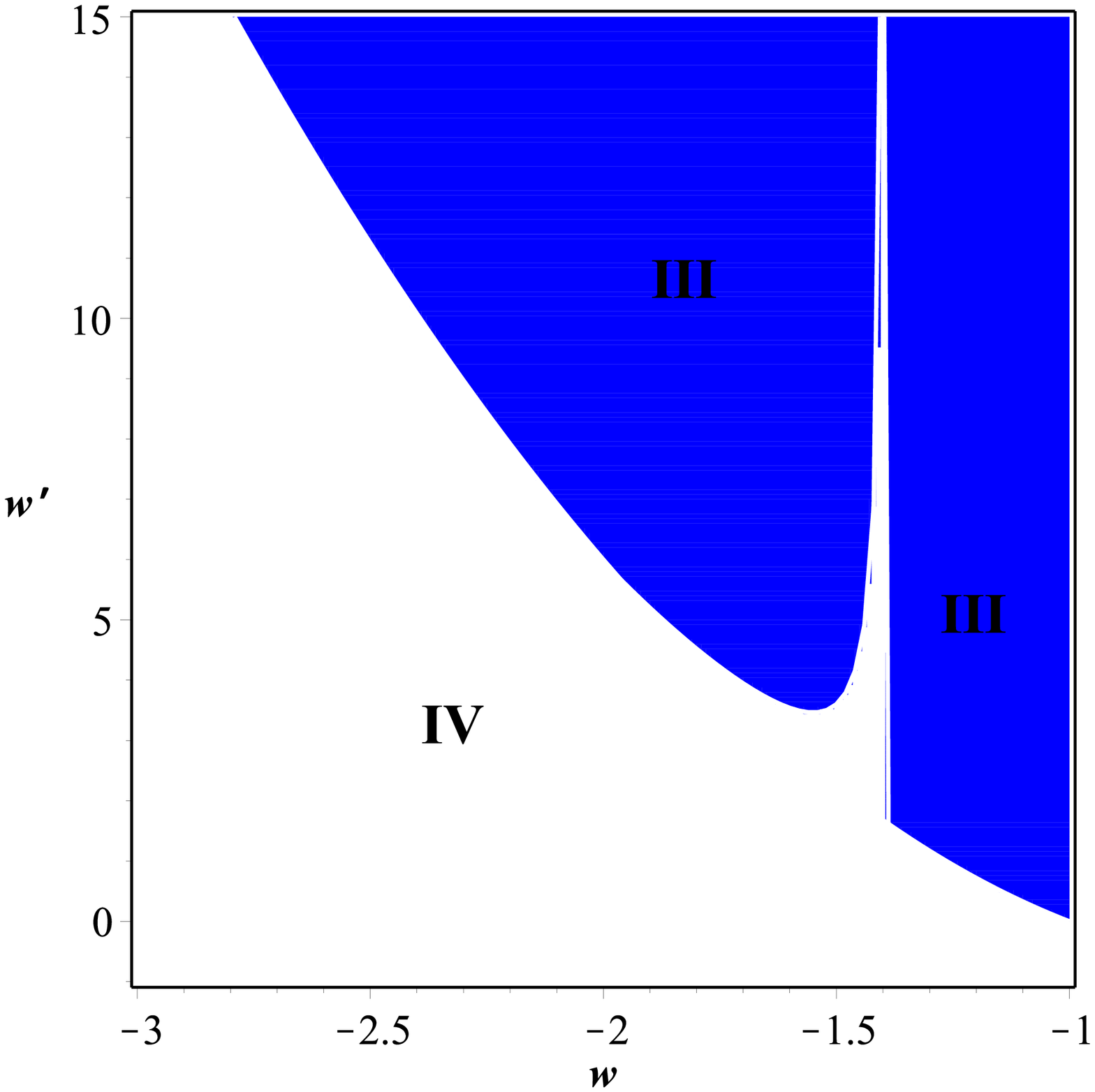}}
\caption{\label{fig3} Bounds on $\omega_{\varphi}'$ as a function of $\omega_{\varphi}$ in $\omega_{\varphi}'-\omega_{\varphi}$ phase plane. Left panel: $\omega_{\varphi}'-\omega_{\varphi}$ phase plane for quintessence field with $\lambda=2$, $\alpha=-2.2$, $x_{1}=0.29$ and $\Omega_{\varphi}=0.69$. Right panel: $\omega_{\varphi}'-\omega_{\varphi}$ phase plane for phantom field with $\lambda=0.5$, $\alpha=-3.5$, $x_{1}=0.31$ and $\Omega_{\varphi}=0.69$.}
\end{figure*}

In terms of the sign of the sound speed, the phase plane is divided into the following four regions. $c_{a}^{2}>0$ is the necessary for the stability of the solutions. As we see in figure \ref{fig3}, the stable regions of the solutions in this case are the regions $I$ and $III$. The
region $I$ belongs to a quintessence phase, while region $III$ is a phantom phase.

\begin{equation}
\left\{
  \begin{array}{ll}
    \omega_{\varphi}>-1, & \hbox{$\omega\acute{}<3\omega_{\varphi}(1+\omega_{\varphi})$}~~\Rightarrow~~c_{a}^2>0~~~(I)\\
    \omega_{\varphi}>-1, & \hbox{$\omega\acute{}>3\omega_{\varphi}(1+\omega_{\varphi})$}~~\Rightarrow~~c_{a}^2<0~~~(II)\\
    \omega_{\varphi}<-1, & \hbox{$\omega\acute{}>3\omega_{\varphi}(1+\omega_{\varphi})$}~~\Rightarrow~~c_{a}^2>0~~~(III)\\
    \omega_{\varphi}<-1, & \hbox{$\omega\acute{}<3\omega_{\varphi}(1+\omega_{\varphi})$}~~\Rightarrow~~c_{a}^2<0~~~(IV)
      \end{array}
\right.
\end{equation}

\section{Statefinder Diagnostic}

Nowadays there are a lot of dark energy models. In 2003
Sahni et al. \cite{Sahni1} have proposed a new pair of parameters $\{r, s\}$, called statefinder
parameters to distinguish between different types of dark energy models.
By using the second and third derivatives of the scale factor these parameters can be achieved. The second
derivative of the expansion factor gives the deceleration parameter. In spatially flat universe it takes the following form
\begin{equation}\label{q}
q=-\frac{\ddot{a}}{aH^2}=-(1+\frac{\dot{H}}{H^2}).
\end{equation}
These statefinder diagnostic pair of parameters, $\{r, s\}$, are defined as

\begin{equation}\label{r}
r=\frac{\dot{}\dot{a}\dot{}}{aH^3}=\frac{\ddot{H}}{H^3}-3q-2,
\end{equation}
\begin{equation}\label{s}
s=\frac{r-1}{3(q-\frac{1}{2})}.
\end{equation}

The statefinder diagnostic tool completely depends on the scale factor.
Describing the spacetime by a metric implies that the
statefinder is a “geometrical” diagnostic tools. From the statefinder diagnostic's point of view,
there are various dark energy models which their evolutionary trajectories in $\{r, s\}$ (or equivalently in $\{r, q\}$ or $\{s, q\}$) plane are different.
So, the statefinder diagnostic tool has an important role in distinguishing between alternative dark energy models.
Besides, the statefinder parameters are useful tools to study the
expansion history of the universe by using higher derivatives of the
scale factor ($\dot{}\dot{a}\dot{}$). Currently the concordance model for dark energy scenario is the $\Lambda CDM $ model which corresponds to fixed point in the $r-s$ phase diagram with $\{r, s\}_{\Lambda CDM} = \{1,0\}$ (or equivalently $\{r, q\}_{\Lambda CDM} = \{1,-1\}$ or $\{s, q\}_{\Lambda CDM} = \{0,-1\}$ ). By describing the trajectories in the $r-s$ phase plane, we
can identify discrepancy of the models from the $\Lambda CDM$ scenario \cite{Sahni2}. The equation (\ref{r}) can be rewritten as

\begin{equation}\label{r1}
r=\frac{d}{dN}\bigg[\frac{\dot{H}}{H^2}\bigg]+2\bigg[\frac{\dot{H}}{H^2}\bigg]^2+3\bigg[\frac{\dot{H}}{H^2}\bigg]+1.
\end{equation}
As we have seen previously, for both quintessence and phantom fields which their derivatives are non-minimally coupled to curvature, there are attractor solutions. By taking derivative from Eq. (\ref{Hdot}), Eq. (\ref{r1}) can be written as follows

\begin{widetext}
\begin{eqnarray}\label{r2}
r=\frac{\dot{H}}{H^2}\Bigg[\frac{\frac{2}{3}(x_1'x_4+x_4'x_1)x_1x_4-\frac{4}{9}\Big((2x_1'x_1x_4^4+4x_1^2x_4'x_4^3)(\epsilon+\frac{x_4^2}{3})
-(\frac{2}{3}x_4'x_4)x_1^2x_4^4\Big)\Big(\epsilon+\frac{x_4^2}{3}\Big)^{-2}}
{\Big(1-\frac{x_1^2x_4^2}{3}+\frac{4x_1^2x_4^4}{9\epsilon+3x_4^2}\Big)}\Bigg]-\nonumber\\
\frac{6x_1'x_1+3\gamma x_3'x_3+2x_1'x_1x_4^2+2x_4'x_4x_1^2}{1-\frac{x_1^2x_4^2}{3}+\frac{4x_1^2x_4^4}{9\epsilon+3x_4^2}}
+\frac{\frac{2}{3}x_4'x_4(\frac{\sqrt{6}}{3}\alpha x_1 x_4^2-\frac{\sqrt{6}}{3}\lambda x_1x_2^2x_4^2+2\epsilon x_1^2x_4^2+\frac{2}{3}x_4'^4x_1^2)}{\Big(\epsilon+\frac{x_4^2}{3}\Big)^2(1-\frac{x_1^2x_4^2}{3}+\frac{4x_1^2x_4^4}
{9\epsilon+3x_4^2})}\nonumber\\
-\frac{\frac{\sqrt{6}}{3}\alpha(x_1x_4^2x_3^2+2x_1x'_{4}x_4x_3^2+x_1x_4^2x_3x'_3)
-\frac{\sqrt{6}}{3}\lambda(x_1'x_2^2x_4^2+2x_1x_2'x_2x_4^2+2x_1x_2^2x_4'x_4)}
{(\epsilon+\frac{x_4^2}{3})(1-\frac{x_1^2x_4^2}{3})+\frac{4x_1^2x_4^4}{9\epsilon+3x_4^2}}\nonumber\\
\frac{-4\epsilon x_1'x_1x_4^2-
4\epsilon x_1^2x_4'x_4-\frac{4}{3}x_1'x_1x_4^4-\frac{8}{3}x_1^2x_4^3}{(\epsilon+\frac{x_4^2}{3})(1-\frac{x_1^2x_4^2}{3})+\frac{4x_1^2x_4^4}
{9\epsilon+3x_4^2}}+2\bigg[\frac{\dot{H}}{H^2}\bigg]^2+3\frac{\dot{H}}{H^2}+1.
\end{eqnarray}
\end{widetext}

where a prime shows derivative with respect to $N=\ln a(t)$ and $x_2',\, x_3',\,  x_4'$ can be replaced by
equations (\ref{x'2})-(\ref{x'4}) respectively. Note also that this relation holds for both quintessence and phantom field due to existence of $\epsilon=\pm1$. \\
Also by substituting equations
(\ref{r2}), (\ref{q}) and (\ref{Hdot}) into equation (\ref{s}), the  parameter $s$ can be derived. For the sake of economy we do not present this lengthy equation explicitly here.

\begin{figure*}
\flushleft\leftskip0em{
\includegraphics[width=.45\textwidth,origin=c,angle=0]{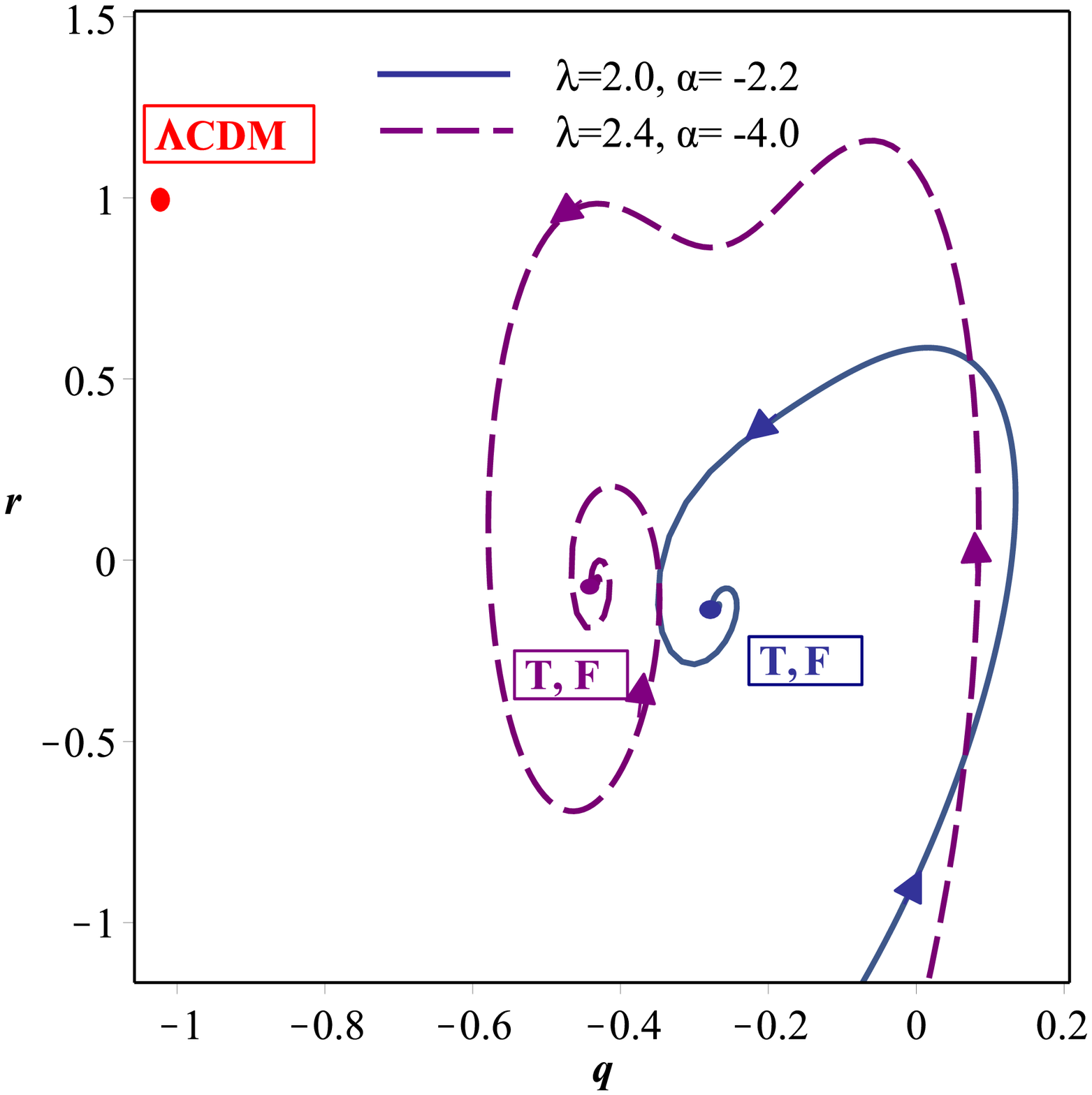}
\hspace{0.5cm}
\includegraphics[width=.45\textwidth,origin=c,angle=0]{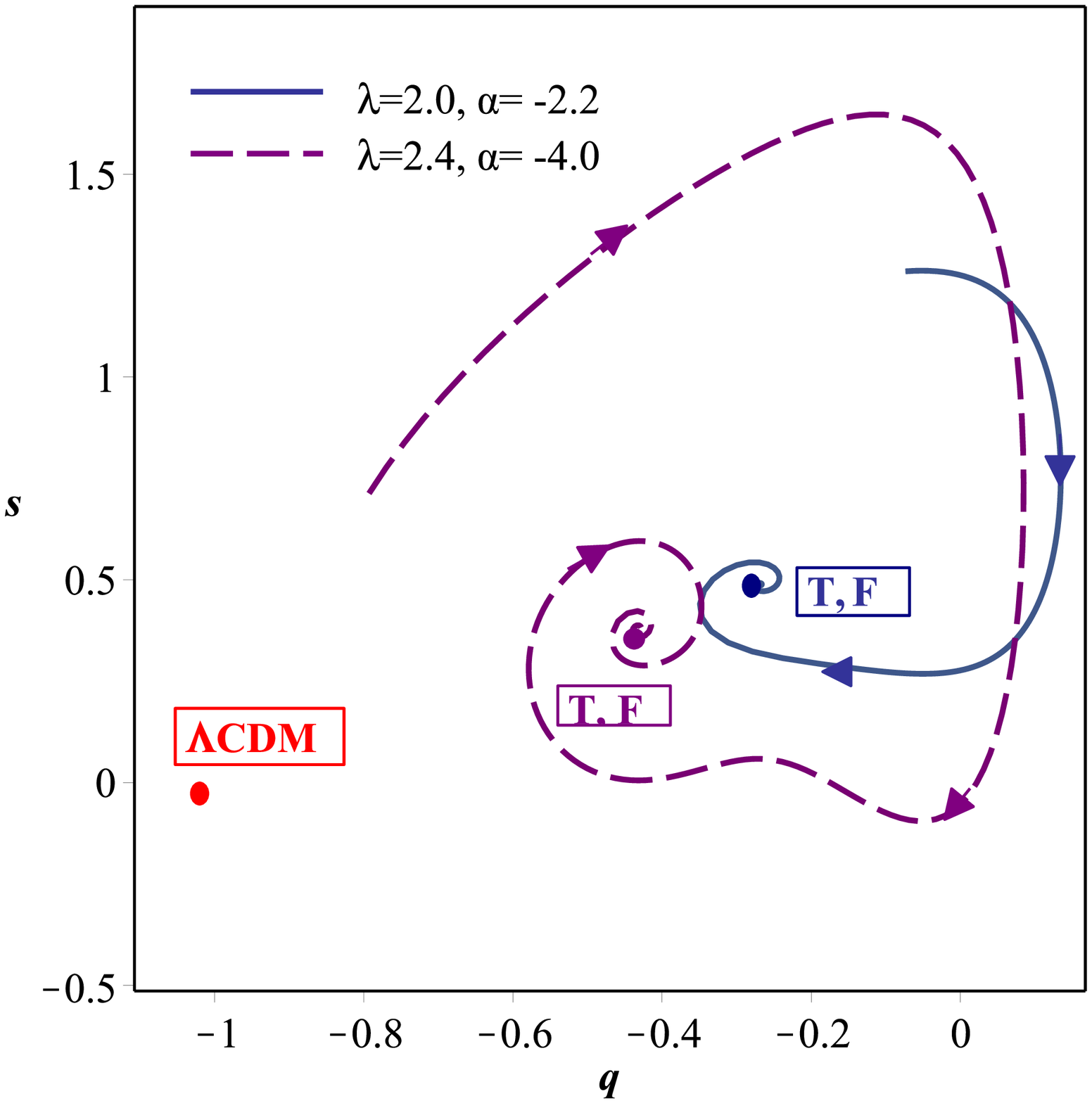}
\hspace{0.5cm}
\includegraphics[width=.45\textwidth,origin=c,angle=0]{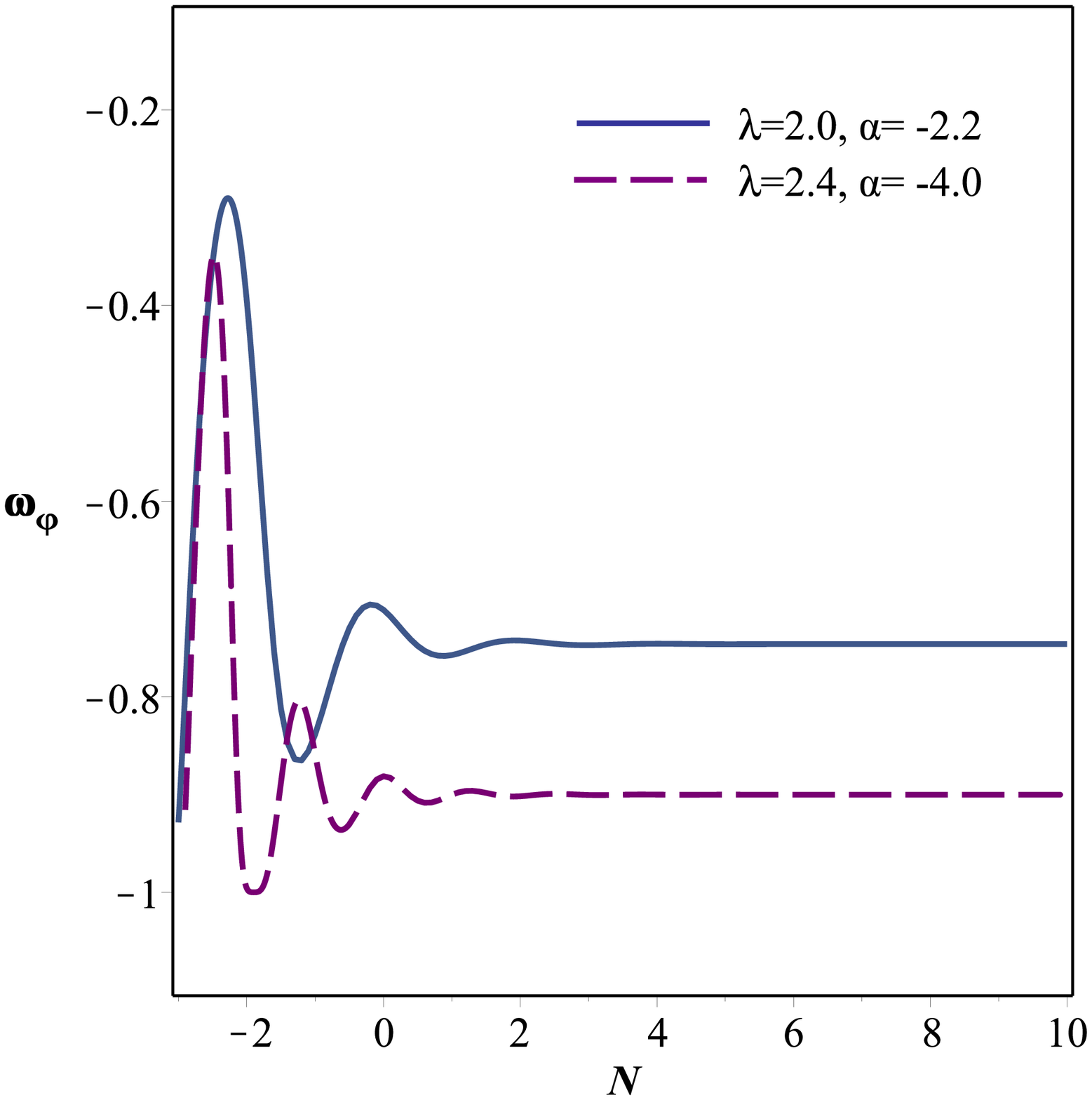}}
\caption{\label{fig4} Trajectories of the stable nodes $F_{\pm\mp}$, in $\{q, r\}$ phase
plane with initial values $x_{2}=0.78$, $x_{3}=0.55$ and $x_{4}=0$ for blue
solid line, and $x_{1}=0.81$, $x_{3}=0.55$ and $x_{4}=0$ for purple dashed line
with the specified values of $\lambda$ and $\alpha$ (upper left
panel). Point $T$ shows the late time values of $\{q, r\}$ in this
model. Point $F$ is the stable state of $\{q, r\}$ in future. The
value of the statefinder $\{q, r\}$ in the $\Lambda CDM$ scenario is
shown by $\Lambda CDM$ point. The upper right panel shows the
trajectories of the stable nodes $F_{\pm\mp}$ in $\{q, s\}$ phase plane
with the same initial values as for the left panel. The lower panel
is devoted to effective equation of
state parameter versus the cosmic time for critical
points $F_{\pm\mp}$ with initial values $x_{3}=0.55$, $x_{2}=0.78$ and $x_{4}=0$ for blue solid line and
$x_{3}=0.55$, $x_{2}=0.81$ and $x_{4}=0$ for purple dashed line with
the values of $\lambda$ and $\alpha$ as given in the figure. The lower panel
is devoted to $\omega_{\varphi}$ parameter.}
\end{figure*}

\begin{figure*}
\flushleft\leftskip0em{
\includegraphics[width=.45\textwidth,origin=c,angle=0]{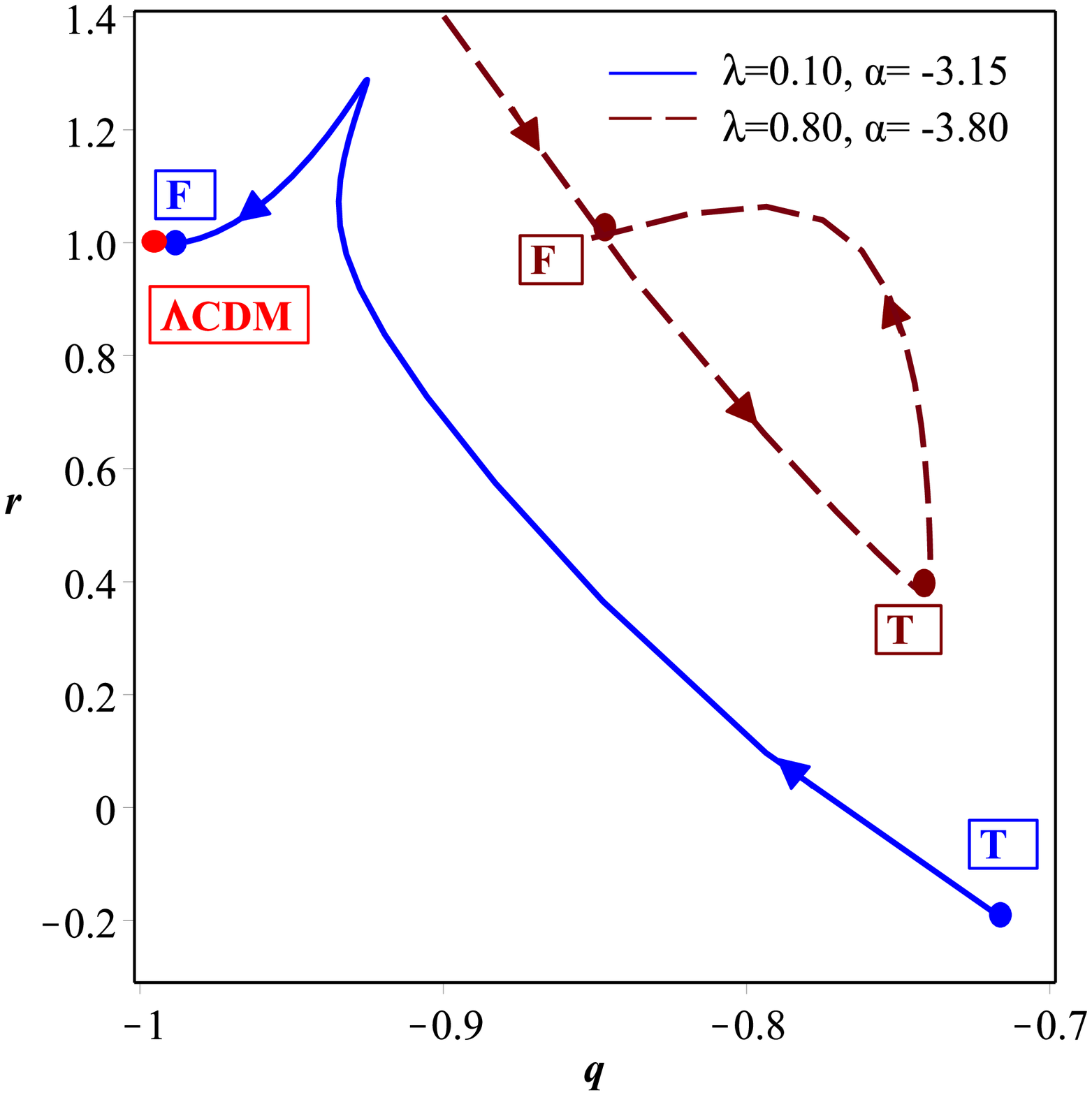}
\hspace{0.5cm}
\includegraphics[width=.45\textwidth,origin=c,angle=0]{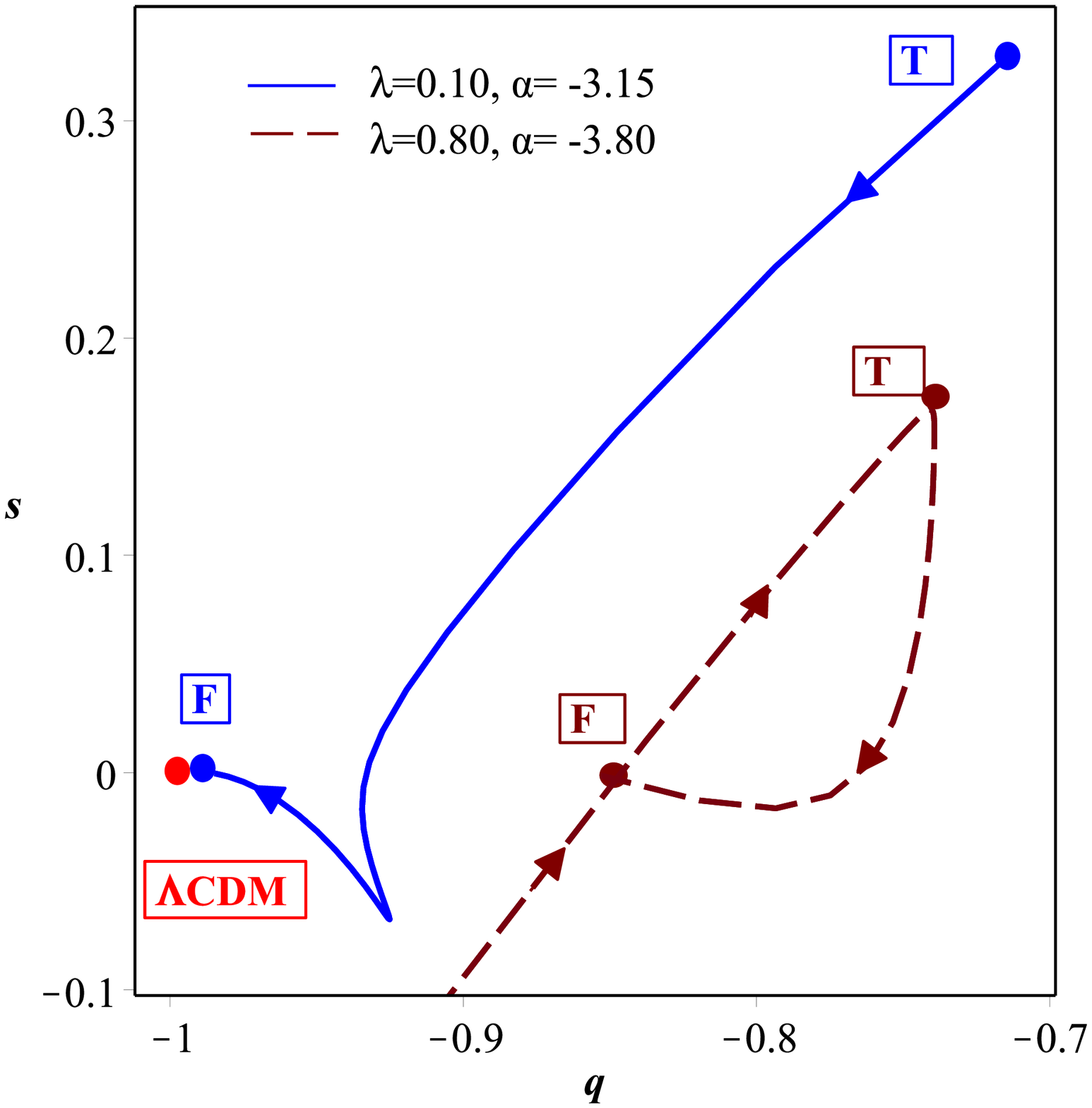}
\hspace{0.5cm}
\includegraphics[width=.45\textwidth,origin=c,angle=0]{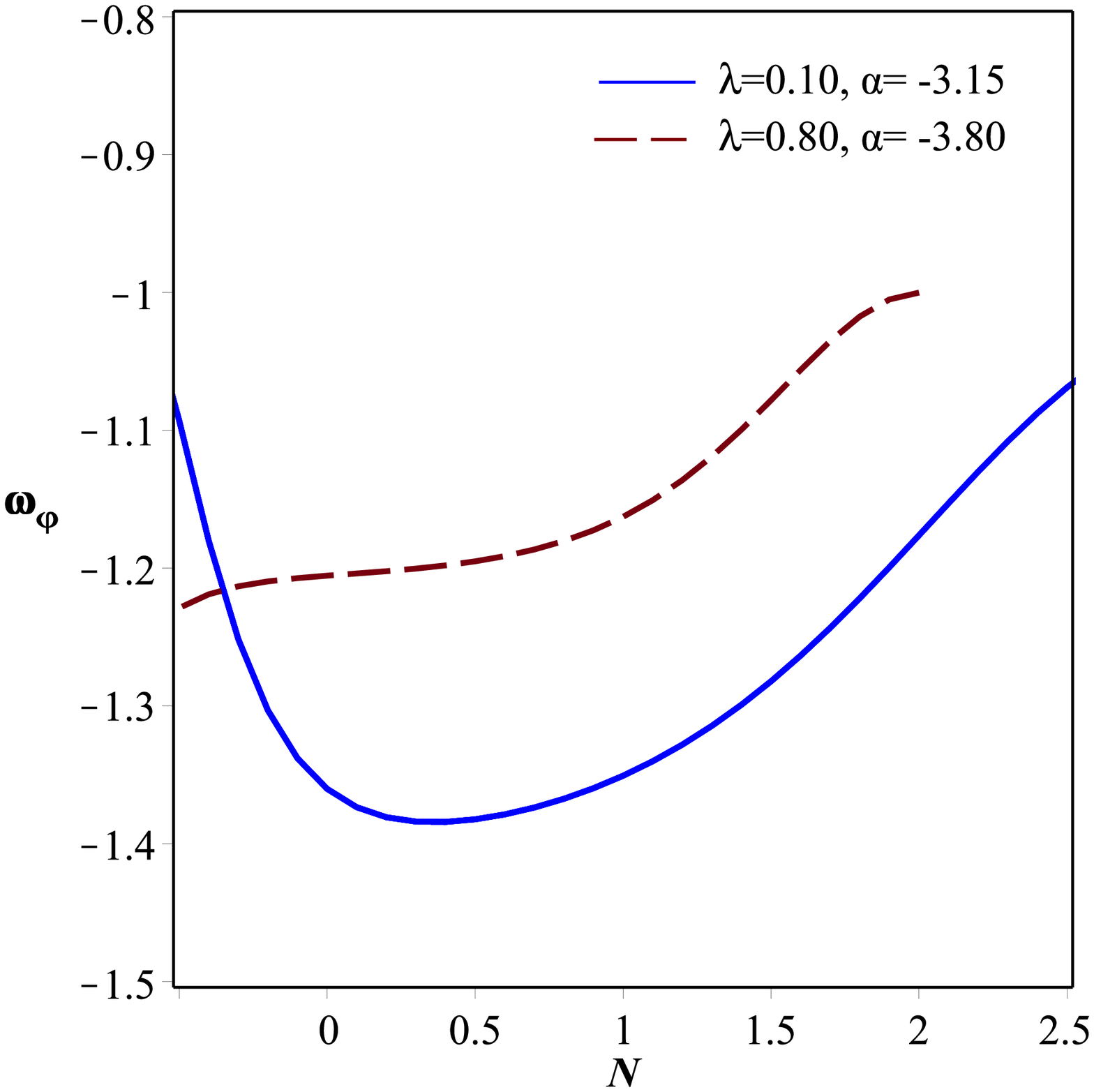}}
\caption{\label{fig5} Trajectories of the stable nodes $F_{\pm\mp}$, in $\{q, r\}$ phase
plane with initial values $x_{2}=0.90$, $x_{3}=0.56$ and $x_{4}=0$ for blue
solid line, and $x_{2}=0.87$, $x_{3}=0.56$ and $x_{4}=0$ for brown dashed line
with the specified values of $\lambda$ and $\alpha$ (upper left
panel). Point $T$ shows the late time values of $\{q, r\}$ in this
model. Point $F$ is the stable state of $\{q, r\}$ in future. The
value of the statefinder $\{q, r\}$ in the $\Lambda CDM$ scenario is
shown by $\Lambda CDM$ point. The upper right panel shows the
trajectories of the stable nodes $F_{\pm\mp}$ in $\{q, s\}$ phase plane
with the same initial values as for the left panel. We see that $\Lambda CDM$ belongs to the cosmic history with a
phantom field with non-minimal derivative coupling. The lower panel
is devoted to $\omega_{\varphi}$ parameter.}
\end{figure*}

Now we study numerically the statefinder
diagnostic for both quintessence and phantom field in this setup. First we consider quintessence field and set $\epsilon=1$ in the above equation. Figure \ref{fig4} illustrates the trajectories of $\{q, r\}$ and $\{q, s\}$ phase plane and $\omega_{\varphi}$ for the critical points $F_{\pm\mp}$ with two different values of $\lambda$ and $\alpha$
parameters. For blue solid line, the late time stable attractor solutions
$F_{\pm\mp}$ indicate that if $\lambda=2$ and $\alpha=-2.2$, the initial values
will be $x_{2}=0.78$ and $x_{3}=0.55$ $(x_{3}^2=\Omega_m=0.31)$ with $x_{4}=0$.
For $\lambda=2.4$ and $\alpha=-4$, the initial values will be $x_{2}=0.81$,
$x_{3}=0.55$ and $x_{4}=0$ for purple dashed line and this is in agreement with
Planck2015 data \cite{planck2015}. The figures indicate that with different values of
$\lambda$ and $\alpha$ parameters the trajectories can evolve
differently. Just by considering larger values of parameter $\alpha$, the trajectories of parameters will approach the $\Lambda CDM$ model. So, the $\Lambda CDM$ is not in the cosmic history of quintessence field with non-minimal derivative coupling at least for small values of parameter $\alpha$.
For phantom field we set $\epsilon=-1$ in equation (\ref{r2}). Figure \ref{fig5} illustrates the trajectories of $\{q, r\}$ and $\{q, s\}$ phase plane and $\omega_{\varphi}$ for critical points $F_{\pm\mp}$ with two different values of the $\lambda$ and $\alpha$
parameters. For blue solid line, the late time stable attractor solutions
$F_{\pm\mp}$ indicate that if $\lambda=0.1$ and $\alpha=-3.15$, the initial values
will be $x_{2}=0.90$ and $x_{3}=0.56$ $(x_{3}^2=\Omega_m=0.31)$ with $x_{4}=0$.
For $\lambda=0.80$ and $\alpha=-3.80$, the initial values will be $x_{2}=0.87$,
$x_{3}=0.56$ and $x_{4}=0$ for brown-dashed line and this is in agreement with
Planck2015 data \cite{planck2015}. Once again, these figures indicate that with different values of
$\lambda$ and $\alpha$ parameters, the trajectories can evolve
differently. Considering smaller values of $\alpha$, the trajectories of parameters will approach the $\Lambda CDM$ model and therefore $\Lambda CDM$ belongs to the cosmic history with a phantom field with non-minimal derivative coupling.

\section{Cosmological Perturbations with Quintessence field}

Now we study cosmological perturbations in our setup with a quintessence field that is coupled non-minimally with dark matter and also its derivatives are coupled to the background curvature. We investigate the analytical solution of matter perturbations in this coupled scenario and compare the results with matter perturbation solutions without interaction between the dark sectors.
In scalar perturbations scenario a FRW perturbed background metric has the
following form (we do not study vector and tensor perturbations):
\begin{eqnarray}
ds^2=-(1+2A)dt^2+2a\partial_iBdx^idt+\nonumber\\
a^2\Big[(1+2\psi)\delta_{ij}+2\partial_{ij}E\Big]dx^idx^j,
\end{eqnarray}
where $\partial_i$ stands $\partial/\partial x^i$, the spatial partial derivative.
Now we define cosmological gauge-invariant variables. The scalar perturbations transform under gauge transformations $t\rightarrow t+\delta t$ and $x\rightarrow x^i=x^i+\delta^{ij}\partial_j\delta x$ as \cite{Bardeen, Mukhanov}

\begin{eqnarray}
&&A\rightarrow A-\dot{\delta}t,~~~~~B\rightarrow B-a^{-1}\delta t+a\dot{\delta}x,\nonumber\\
&&\psi\rightarrow\psi-H\delta t,~~~~~E\rightarrow E-\delta x,
\end{eqnarray}
Also the field perturbation transformation is

\begin{equation}
\delta\varphi\rightarrow\delta\varphi-\dot{\varphi}\delta t,
\end{equation}
It is convenient to consider Bardeen’s or gauge invariant potentials which firstly were introduced in \cite{Bardeen}. There are, in addition, other gauge
invariant variables, (see for instance \cite{Hwang,Hwang1}), but here we just consider Bardeen's potentials as follow:

\begin{equation}\label{Bardeen1}
\Phi\equiv A-\frac{d}{dt}\Bigg[a^2\Big(\dot{E}+\frac{B}{a}\Big)\Bigg],
\end{equation}
\begin{equation}\label{Bardeen2}
\Psi\equiv -\psi+a^2H\Big(\dot{E}+\frac{B}{a}\Big).
\end{equation}
It is useful to consider the following decomposition of energy-momentum tensor for perturbation calculations

\begin{eqnarray}
T_0^0=&&-(\rho+\delta\rho),~~T^0_{\alpha}=-(\rho+P)\nu,_{\alpha},\nonumber\\
&&T^{\alpha}_{\beta}=(p+\delta p)\delta^{\alpha}_{\beta}+\Pi^{\alpha}_{\beta},
\end{eqnarray}
where $\Pi^{\alpha}_{\beta}$ is a traceless anisotropic stress.
We suppose the Newtonian (longitudinal or shear-free) gauge in which $B=E=0$. At the linear order for action (\ref{action}), perturbed Einstein equations, $\delta G^{\mu}_{\nu}=\delta T^{\mu}_{\nu}$, and also covariant divergence of the perturbed tensor
$\delta\nabla_{\mu}T^{\mu}_{\nu}=\delta(Q\rho_m\dot{\varphi})$ for $\nu=0,i$ can be derived respectively as

\begin{eqnarray}\label{pe1}
3H&&(\dot{\psi}-HA)-\frac{\nabla^2\psi}{a^2}=\frac{1}{2}\bigg[\dot{\varphi}\delta\dot{\varphi}-\dot{\varphi}^{2}A+\frac{\partial V}{\partial \varphi}\delta\varphi+\nonumber\\
&&\eta\Big(-9H\dot{\varphi}^2
\dot{\psi}+9H^2\dot{\varphi}\delta\dot{\varphi}+\frac{2}{a^2}H\dot{\varphi}\nabla^2\delta\varphi\nonumber\\
&&-\frac{\dot{\varphi}^2}{a^2}\nabla^2\psi
+F'(\varphi)\delta\varphi\rho_{m}+F(\varphi)\delta\rho_{m}\Big)\bigg],
\end{eqnarray}

\begin{eqnarray}\label{pe2}
HA-\dot{\psi}&&=\frac{1}{2}\Big[\dot{\varphi}\delta\varphi+\eta(3H\dot{\varphi}^2A-\dot{\varphi}^2\dot{\psi}+3H^2\dot{\varphi}\delta\varphi)\nonumber\\
&&+2H\dot{\varphi}\delta\dot{\varphi}+aF(\varphi)(\rho_{m}+p_{m})\nu_{m}\Big],
\end{eqnarray}

\begin{equation}\label{pe3}
A=-\psi+\frac{1}{2}\eta\Big[(A-\psi)\dot{\varphi}^2+2\ddot{\varphi}\delta\varphi+2H\dot{\varphi}\delta\varphi\Big],
\end{equation}

\begin{eqnarray}\label{pe4}
&&3H^2A+2\dot{H}A+H\dot{A}+\frac{1}{3a^2}\nabla^2A-3H\dot{\psi}-\ddot{\psi}=\nonumber\\
&&-\frac{1}{3a^2}\nabla^2\psi+\frac{1}{2}\Big[\dot{\varphi}\delta\dot{\varphi}-\dot{\varphi}^2A-\frac{\partial V(\varphi)}{\partial\varphi}\delta\varphi+\nonumber\\
&&\eta\Big[(6H^2+4\dot{H})\dot{\varphi}^2A+8H\dot{\varphi}\ddot{\varphi}A+3H\dot{\varphi}^2\dot{A}+\nonumber\\
&&\frac{1}{3a^2}\dot{\varphi}^2\nabla^2A-3H\dot{\varphi}^2\dot{\psi}-3H\dot{\varphi}\ddot{\varphi}\dot{\psi}-\dot{\varphi}^2\ddot{\psi}
-\frac{1}{3a^2}\dot{\varphi}^2\nabla^2\psi\nonumber\\
&&-3H^2\dot{\varphi}\delta\dot{\varphi}-2\dot{H}\dot{\varphi}\delta\dot{\varphi}-2H\ddot{\varphi}\delta\dot{\varphi}
-2H\dot{\varphi}\delta\ddot{\varphi}+\nonumber\\
&&\frac{2}{3a^2}(\ddot{\varphi}+H\dot{\varphi}+\nabla^2\delta\varphi)\Big]
+F'(\varphi)\delta\varphi p_{m}+F(\varphi)\delta p_{m}\Big],
\end{eqnarray}

\begin{eqnarray}\label{deltam}
&&\delta(F(\varphi)\rho_{m}\dot{)}+3H\delta\Big(F(\varphi)(\rho_{m}+p_{m})\Big)=\nonumber\\
&&~~~~~F(\varphi)(\rho_{m}+p_{m})\Big(-3\dot{\psi}+\frac{\nabla^2}{a}\nu_{m}\Big)\nonumber\\
&&+\delta F'(\varphi)\rho_{m}\dot{\varphi}+F'(\varphi)\delta\rho_{m}\dot{\varphi}+F'(\varphi)\rho_{m}\delta\dot{\varphi},
\end{eqnarray}

\begin{eqnarray}\label{vdot}
\dot{\nu}_{m}+\Big[(1-3\omega_{m})H+\frac{F'(\varphi)}{F(\varphi)}\dot{\varphi}\Big]\nu_{m}=\nonumber\\
\frac{1}{a}\Big[A+\frac{\omega_{m}}{1+\omega_{m}}\delta_m
+\frac{F'(\varphi)}{F(\varphi)}\frac{\delta\varphi}{1+\omega_m}\Big].
\end{eqnarray}
where a prime denotes derivative with respect to $\varphi$.
We investigate the evolution of perturbations on sub-Hubble scales. For this end, we are interested in to calculate $\delta_m\equiv\delta(F(\varphi)\rho_m)/F(\varphi)\rho_m$ in order to derive matter perturbations. Note that large scale
galaxy clustering observations induce constraints on the dark energy, and in this case this parameter plays an important role. In which follows, we consider $\omega_m$ as a constant. By adopting the following Fourier transformations

\begin{equation}\label{Fourier1}
\varphi(x,t)\rightarrow e^{ik.r}\varphi(t),
\end{equation}

\begin{equation}\label{Fourier2}
\nabla\varphi(x,t)\rightarrow ie^{ik.r}k\varphi(t),
\end{equation}

\begin{equation}\label{Fourier3}
\nabla^2\varphi(x,t)\rightarrow -e^{ik.r}k^2\varphi(t)\,,
\end{equation}
equations (\ref{deltam}) and (\ref{vdot}) can be read as follows

\begin{equation}\label{deltamm}
\dot{\delta}_{m}=(1+\omega_{m})(3\dot{\Psi}-\frac{k^2}{a}\nu_{m})+\frac{\delta F'(\varphi)}{F(\varphi)}\dot{\varphi}+
\frac{F'(\varphi)}{F(\varphi)}\delta\dot{\varphi},
\end{equation}

\begin{eqnarray}
\dot{\nu}_{m}+\Big[(1-3\omega_{m})H+\frac{F'(\varphi)}{F(\varphi)}\dot{\varphi}\Big]\nu_{m}=\nonumber\\
\frac{1}{a}\Big[\Phi+\frac{\omega_{m}}{1+\omega_{m}}\delta_m
+\frac{F'(\varphi)}{F(\varphi)}\frac{\delta\varphi}{1+\omega_m}\Big].
\end{eqnarray}

Notice that, since the equations are linear, the Fourier modes $e^{ik.r}$ can be easily dropped out. Taking a derivative from Eq. (\ref{deltamm}) and eliminating $\nu_m$ from the two above equations, we obtain

\begin{eqnarray}\label{deltammm}
&&~~~~~~~~~~\ddot{\delta}_{m}+\bigg[(2-3\omega_m)H+C\bigg]\dot{\delta}_m+\nonumber\\
&&~~~~~\frac{k^2}{a^2}\omega_m\delta_m+\frac{k^2}{a^2}
\bigg[\Phi+\frac{E}{1+\omega_m}\bigg](1+\omega_m)=\nonumber\\
&&3(1+\omega_m)\Bigg[\ddot{\Psi}+\bigg[(2-3\omega_m)H+C\bigg]
\bigg[\dot{\Psi}+\frac{D+B}{3(1+\omega_m)}\bigg]\Bigg]\nonumber\\
&&~~~~~~~~~~~~~~~~~~~~+\dot{D}+\dot{B}.
\end{eqnarray}
where

\begin{eqnarray}
&&D=\frac{\delta F'(\varphi)}{F(\varphi)}\dot{\varphi}, \,\,B=\frac{F'(\varphi)}{F(\varphi)}\delta\dot{\varphi}\nonumber\\
&&C=\frac{F'(\varphi)}{F(\varphi)}\dot{\varphi},\,\,\,E=\frac{F'(\varphi)}{F(\varphi)}\delta\varphi,\nonumber
\end{eqnarray}

and
\begin{eqnarray}
&&\dot{D}=\frac{F'''(\varphi)\delta\varphi}{F(\varphi)}\dot{\varphi}^2+\frac{F''(\varphi)}{F(\varphi)}\delta\dot{\varphi}\dot{\varphi}\nonumber\\
+&&\frac{F''(\varphi)\delta\varphi}{F(\varphi)}\ddot{\varphi}-\frac{F''(\varphi)F'(\varphi)\delta\varphi\dot{\varphi}^2}{F^{2}(\varphi)},\nonumber
\end{eqnarray}

\begin{eqnarray}
\dot{B}=&&\frac{F''(\varphi)}{F(\varphi)}\delta\dot{\varphi}\dot{\varphi}+\frac{F'(\varphi)}{F(\varphi)}\delta\ddot{\varphi}-
\frac{F'^2(\varphi)\dot{\varphi}}{F^2(\varphi)}\delta\dot{\varphi}.\nonumber
\end{eqnarray}

We consider non-relativistic matter $(\omega_m=0)$ on scales much smaller than the Hubble radius $(k\gg aH)$. In this case, Eq. (\ref{deltammm}) takes the following form

\begin{equation}
\ddot{\delta}_{m}+\bigg[2H+\frac{F'(\varphi)}{F(\varphi)}\dot{\varphi}\bigg]\dot{\delta}_m+\frac{k^2}{a^2}
\bigg[\Phi+\frac{F'(\varphi)}{F(\varphi)}\delta\varphi\bigg]=0,
\end{equation}
By considering $F'(\varphi)=\alpha F(\varphi)$, $\dot{\varphi}$ in terms of dimensionless parameter from (\ref{di}) and $\delta\varphi$ from perturbed Klein-Gordon equation, we get

\begin{equation}
\ddot{\delta}_{m}+\Big(2H+\sqrt{6}\alpha Hx_1\Big)\dot{\delta}_m-
\frac{3}{2}H^2\Omega_m\Big(1+2\alpha^2\Big)\delta_m=0.
\end{equation}
Using the relation $d/dN=(1/H)(d/dt)$ we obtain

\begin{equation}\label{variationn}
\frac{d^2\delta_m}{dN^2}+\bigg(2+\frac{1}{H}\frac{dH}{dN}+\sqrt{6}\alpha x_1\bigg)\frac{d\delta_m}{dN}
-\frac{3}{2}\Omega_m\Big(1+2\alpha^2\Big)\delta_m=0.
\end{equation}
With this result, we study dynamics of generated perturbations in this setup.

\subsection{Perturbations in the matter domination era}

By choosing transient regime corresponding to the critical points $C_{\pm}$ in table \ref{tab:1}, (that is $\alpha^2<\frac{3}{2}$ and $-\alpha^2+\lambda\alpha<\frac{3}{2}$), we have

\begin{eqnarray}
x_1=\pm\sqrt{\frac{2}{3}}\alpha,~~~\Omega_\varphi=\frac{2}{3}\alpha^2,~~~\omega_{totc}=\frac{2}{3}\alpha^2
\end{eqnarray}
and using the relation $\Omega_m=1-\Omega_{\varphi}$, we get

\begin{equation}\label{dh/dn}
\frac{1}{H}\frac{dH}{dN}=-\frac{3}{2}(1+\omega_{totc}),
\end{equation}
Then equation (\ref{variationn}) takes the following form

\begin{eqnarray}\label{x--}
\frac{d^2\delta_m}{dN^2}+\Big(\frac{1}{2}-3\alpha^2\Big)&&\frac{d\delta_m}{dN}+
\Big(1+2\alpha^2\Big)\Big(-\frac{3}{2}+\alpha^2\Big)\delta_m=0\nonumber\\
&&for\,\,\, x_1=-\sqrt{\frac{2}{3}}\alpha
\end{eqnarray}

\begin{eqnarray}\label{x++}
\frac{d^2\delta_m}{dN^2}+\Big(\frac{1}{2}+\alpha^2\Big)&&\frac{d\delta_m}{dN}+
\Big(1+2\alpha^2\Big)\Big(-\frac{3}{2}+\alpha^2\Big)\delta_m=0\nonumber\\
&&for\,\,\, x_1=+\sqrt{\frac{2}{3}}\alpha
\end{eqnarray}
Solving these two equations we reach at the general solutions

\begin{equation}\label{anal}
\delta_m=C_+a^{n_+}+C_-a^{n_-}.
\end{equation}
For equation (\ref{x--}) we have

\begin{equation}
n_+=1+\alpha^2,\,\,\,\,\,\,\,n_-=-\frac{3}{2}+\alpha^2
\end{equation}
For comparison, in the minimal case (see for instance \cite{Copeland}), $\delta_{m}\propto a$ in order to have formation of galaxy clustering in matter domination era. It is easy to see that just in $n_{+}$ case, when $\alpha=0$, we reach $\delta_{m}\propto a$. For this reason, we ignore the $n_{-}$ case. For $n_{+}$ case, if we consider interaction between dark energy and dark matter, the growth rate is higher than the minimal case and this rate depends on the value of $\alpha$.

For equation (\ref{x++}) we have
\begin{equation}
n_\pm=-\frac{1}{2}\alpha^2-\frac{1}{4}\pm\frac{1}{4}\sqrt{-28\alpha^4+36\alpha^2+25}
\end{equation}

\begin{figure*}
\flushleft\leftskip0em{
\hspace{0.5cm}
\includegraphics[width=.45\textwidth,origin=c,angle=0]{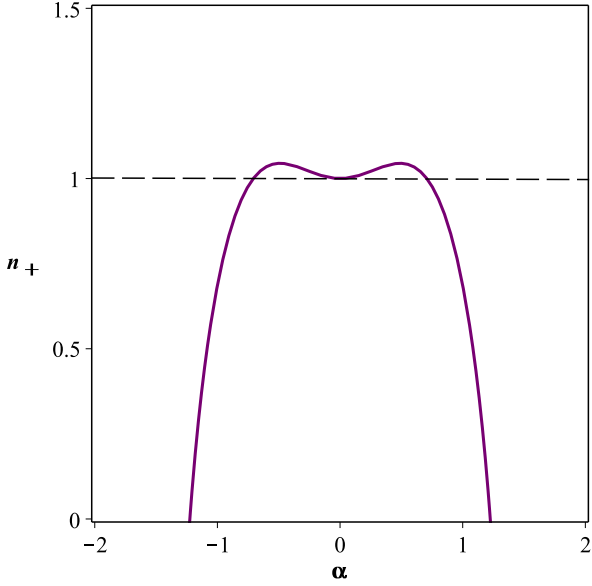}}
\caption{\label{fig6} The growth rate of matter perturbation versus $\alpha$ in the matter domination era. }
\end{figure*}

By the same reason, here we consider just the $n_{+}$ case (with $\alpha=0$, just in $n_+$ case $\delta_{m}\propto a$). However, if we consider interaction between dark energy and dark matter, the growth rate of matter perturbation depends on the value of $\alpha$. This is shown in figure \ref{fig6}. For small values of $\alpha$, it is slightly more than $1$, but for other values of $\alpha$, the growth rate of matter perturbation is less than the value for the minimal case. We note that in the matter domination era the values of $\alpha$ are extremely small, therefore the growth rate of matter perturbation is slightly more than the value for the minimal case.

\subsection{Perturbations in the scaling solution era}

Now we investigate the growth rate of matter perturbations in the scaling solution era. From table \ref{tab:1} for critical points $F_{\pm}$ that have scaling solutions, we have

\begin{eqnarray}
&&~~~~~~~~~~~~~~~x_1=\pm\frac{1}{2}\frac{\sqrt{6}}{\alpha-\lambda},\nonumber\\
&&\Omega_{\varphi}=\frac{\alpha^2-\alpha\lambda+3}{(\alpha-\lambda)^2},~~~\omega_{totc}=\frac{-\alpha}{\alpha-\lambda}
\end{eqnarray}
By replacing the above relations, $\Omega_m=1-\Omega_{\varphi}$ and equation (\ref{dh/dn}) in equation (\ref{variationn}), we get

\begin{eqnarray}\label{sdeltamp}
&&\frac{d^2\delta_m}{dN^2}+\bigg(\frac{1}{2}+\frac{9}{2}(\frac{\alpha}{\alpha-\lambda})\bigg)\frac{d\delta_m}{dN}\nonumber\\
-\frac{3}{2}&&\bigg(\frac{\lambda^2-\alpha\lambda-3}{(\alpha-\lambda)^2}\bigg)\Big(1+2\alpha^2\Big)\delta_m=0\nonumber\\
&&~~~~for\,\,x_1=+\frac{1}{2}\frac{\sqrt{6}}{\alpha-\lambda}
\end{eqnarray}

\begin{eqnarray}\label{sdeltamn}
&&\frac{d^2\delta_m}{dN^2}+\bigg(\frac{1}{2}-\frac{3}{2}(\frac{\alpha}{\alpha-\lambda})\bigg)\frac{d\delta_m}{dN}\nonumber\\
-\frac{3}{2}&&\bigg(\frac{\lambda^2-\alpha\lambda-3}{(\alpha-\lambda)^2}\bigg)\Big(1+2\alpha^2\Big)\delta_m=0\nonumber\\
&&~~~~for\,\,x_1=-\frac{1}{2}\frac{\sqrt{6}}{\alpha-\lambda}
\end{eqnarray}

\begin{figure*}
\flushleft\leftskip0em{
\includegraphics[width=.45\textwidth,origin=c,angle=0]{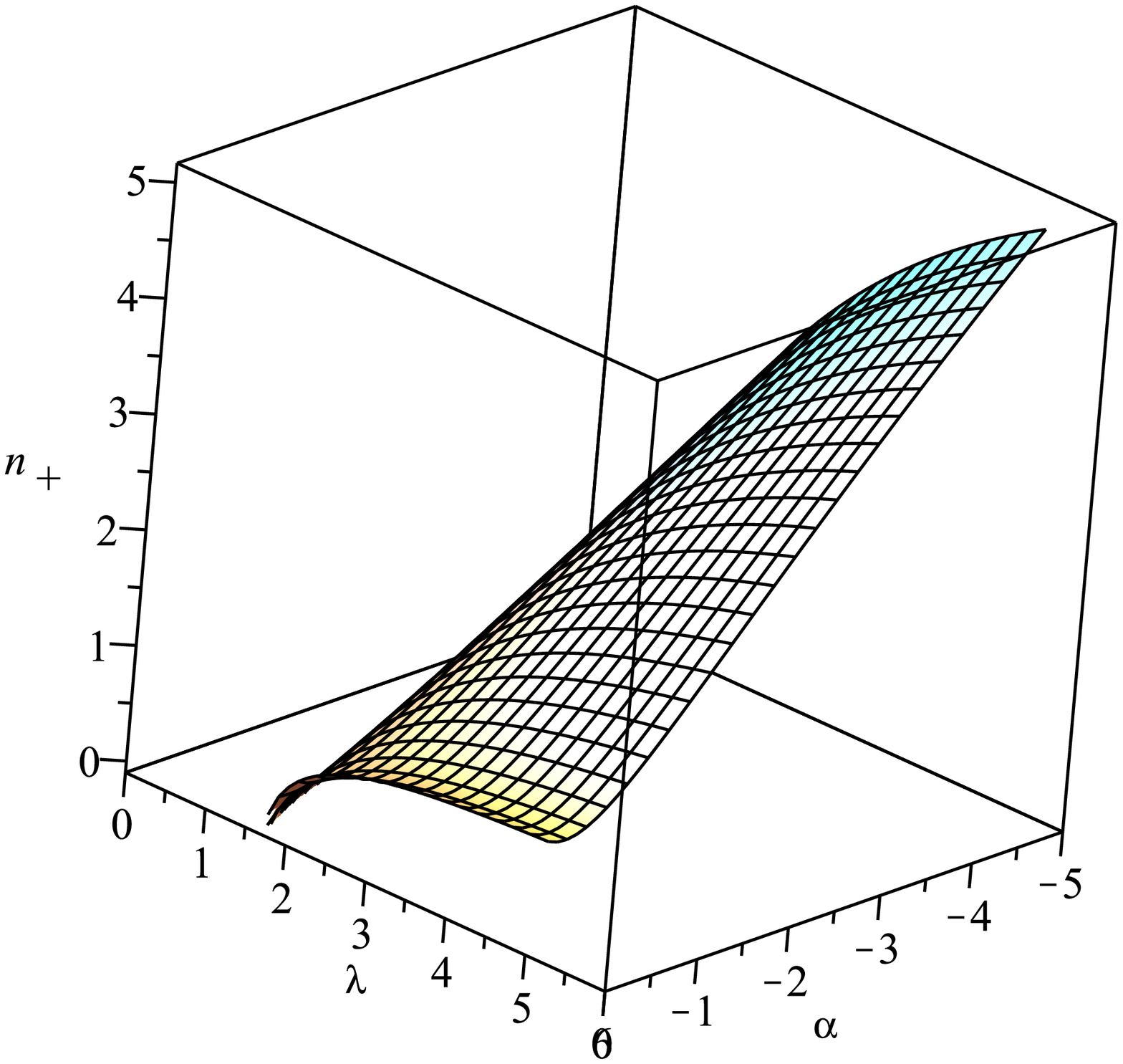}
\hspace{0.5cm}
\includegraphics[width=.45\textwidth,origin=c,angle=0]{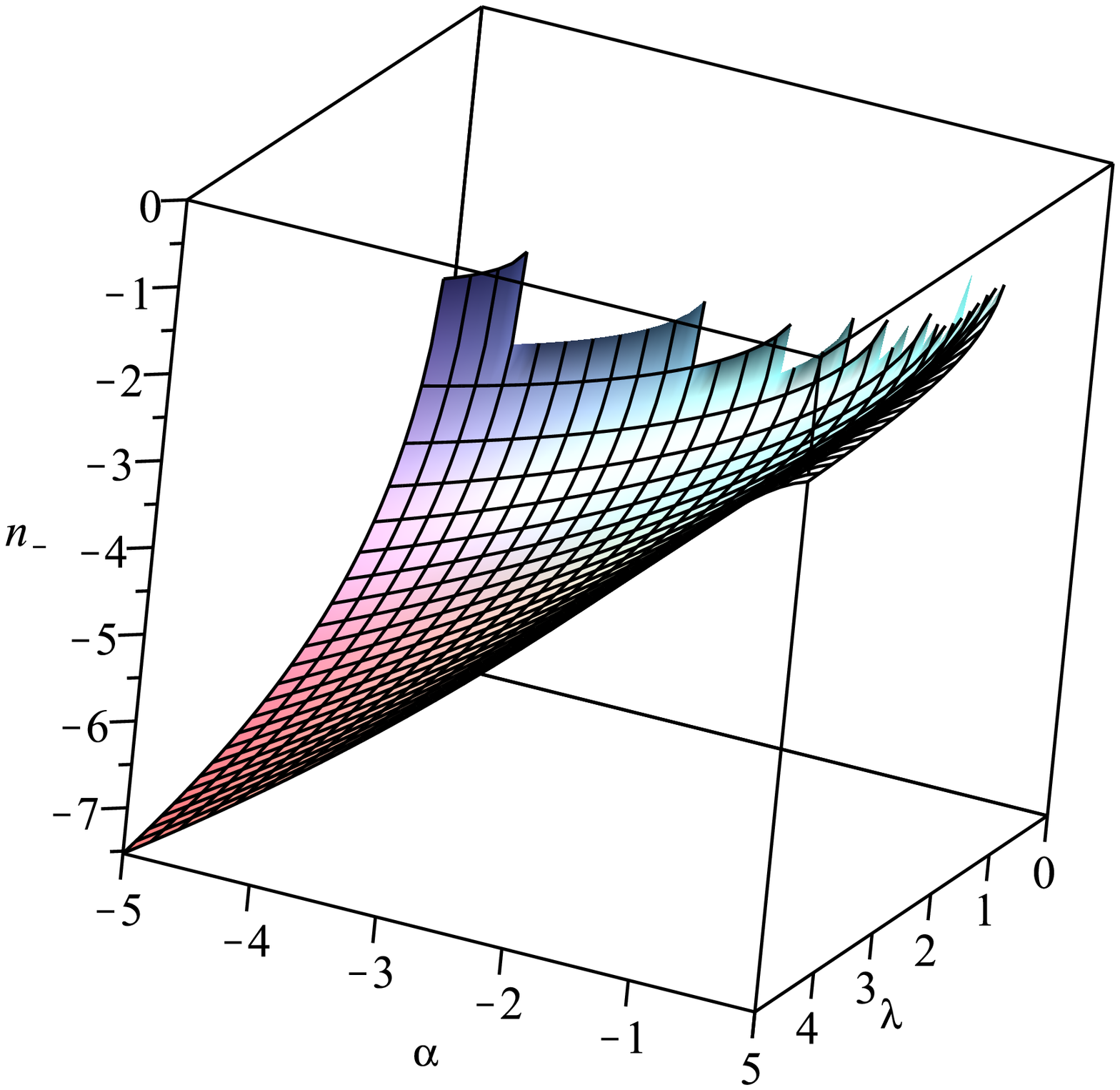}}
\caption{\label{fig7} The growth rate of matter perturbations versus $\alpha$ and $\lambda$ in the present scaling solution era (Eq. (\ref{r1})).}
\end{figure*}

Solving the equation (\ref{sdeltamp}), we obtain the analytical solutions like equation (\ref{anal}) with the following $n_{\pm}$

\begin{eqnarray}\label{r1}
&&n_{\pm}=\frac{1}{4}\Big(\frac{-10\alpha+\lambda}{\alpha-\lambda}\Big)\pm\nonumber\\
\frac{1}{4}&&\bigg(\frac{\sqrt{-48\alpha^3\lambda+48\alpha^2\lambda^2-44\alpha^2-44\alpha\lambda+25\lambda^2-72}}{|\alpha-\lambda|}\bigg)\nonumber\\
\end{eqnarray}
solving the equation (\ref{sdeltamn}) we obtain the analytic solution like equation (\ref{anal}) with following $n_{\pm}$,

\begin{eqnarray}\label{r2}
&&n_{\pm}=\frac{1}{4}\Big(\frac{2\alpha+\lambda}{\alpha-\lambda}\Big)\pm\nonumber\\
\frac{1}{4}&&\bigg(\frac{\sqrt{-48\alpha^3\lambda+48\alpha^2\lambda^2-140\alpha^2-20\alpha\lambda+25\lambda^2-72}}{|\alpha-\lambda|}\bigg)\nonumber\\
\end{eqnarray}
So, in this case the growth rate of matter perturbations depends on two quantities $\alpha$ and $\lambda$. As we see from figure \ref{fig7} for positive $\lambda$ and negative $\alpha$ that we have derived in the previous section, $n_{+}>0$ and $n_{-}<0$. Therefore, we consider just $n_{+}$. For uncoupled dark energy and dark matter $n_+\leq1$ (see \cite{Copeland}), but in our interacting model $n_{+}$ becomes larger than $1$ in both cases.

\section{The Phase Space Analysis }

We consider the case with negative Q (where the energy flows from dark energy to dark matter) and investigate the cosmological status of this model via a dynamical system analysis. We also focus mainly on the role of the non-minimal derivative coupling and the sign of the interaction between the dark sectors. For this purpose,
we use the dimensionless quantities defined in (\ref{di}) to translate the dynamical equations to an autonomous system.
As before, this allows us to investigate evolution of just three variables since the
forth one can be expressed in terms of the other ones. In which follows we
consider $x_1$ as our dependent variable and omit it in our forthcoming calculations.
We rewrite the Friedmann equation (\ref{ac}) and the equation of motion (\ref{motion}) versus the new phase space variables
\begin{eqnarray}\label{Hdot1}
\frac{\dot{H}}{H^2}=&&\frac{1}{1-(\frac{1}{3}-\frac{\frac{4}{9}x_4^2}{\epsilon+\frac{1}{3}x_4^2})x_1^2x_4^2}
\Bigg[-3\epsilon x_1^2-\frac{3}{2}\gamma x_3^2-x_1^2x_4^2+\nonumber\\
&&\frac{(-2\epsilon x_1+\frac{\sqrt{6}}{3}\lambda x_2^2-\frac{\sqrt{6}}{3}\alpha x_3^2-\frac{2}{3}x_1x_4^2)x_1x_4^2}{\epsilon+\frac{1}{3}x_4^2}\Bigg],
\end{eqnarray}

\begin{equation}\label{varphidot1}
\frac{\ddot{\varphi}}{H^2}=\frac{-3\sqrt{6}\epsilon x_1+3\lambda x_2^2-3\alpha x_3^2-\sqrt{6}(\frac{2}{3}\frac{\dot{H}}{H^2}+1)x_1x_4^2}
{\epsilon+\frac{1}{3}x_4^2},
\end{equation}
Considering a new time variable as $N = \ln a(t)$, the equations (\ref{x'2}) and (\ref{x'4}) are valid but equation (\ref{x'3}) changes the sign as follows

\begin{equation}\label{x'33}
x'_3=-\bigg(\frac{3}{2}\gamma-\frac{\sqrt{6}}{2}\alpha x_1+\frac{\dot{H}}{H^2}\bigg)x_3,
\end{equation}

\subsection{The phase space with a quintessence field}

Solving equations (\ref{x'2}), (\ref{x'4}) and (\ref{x'33}) with $\epsilon=+1$ we reach at seven critical points $(A, B, C, D, E, F, G)$ in our system, but the critical point $G$ is not a physically acceptable point, so we just investigate the remaining six points.
The results are summarized in tables \ref{tab:5} and \ref{tab:6}. Now we discuss properties of each critical point separately. In all calculations we consider the condition $\gamma=1$ (a pressureless matter).

\begin{figure*}
\flushleft\leftskip0em{
\includegraphics[width=.45\textwidth,origin=c,angle=0]{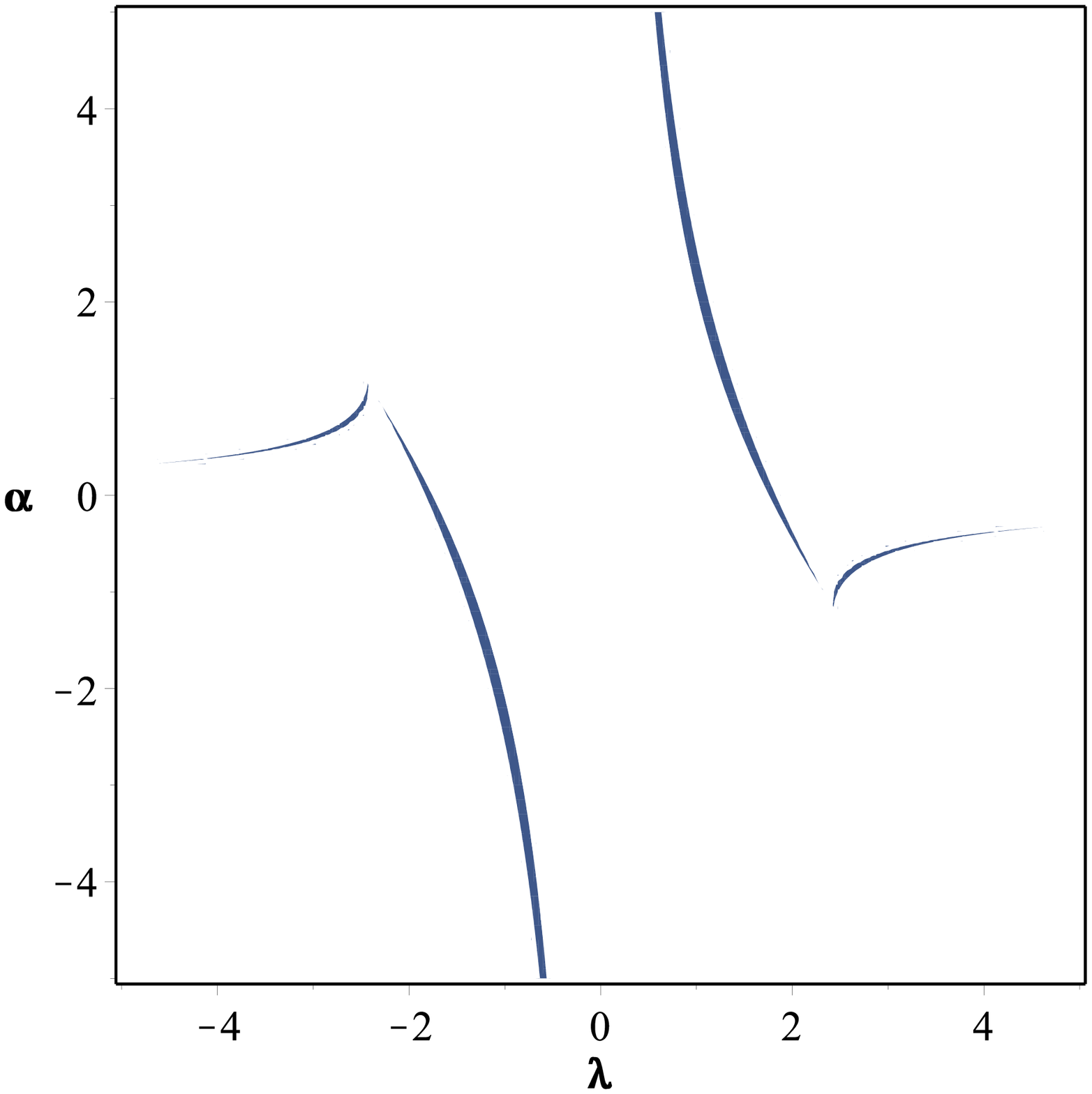}
\hspace{0.5cm}
\includegraphics[width=.45\textwidth,origin=c,angle=0]{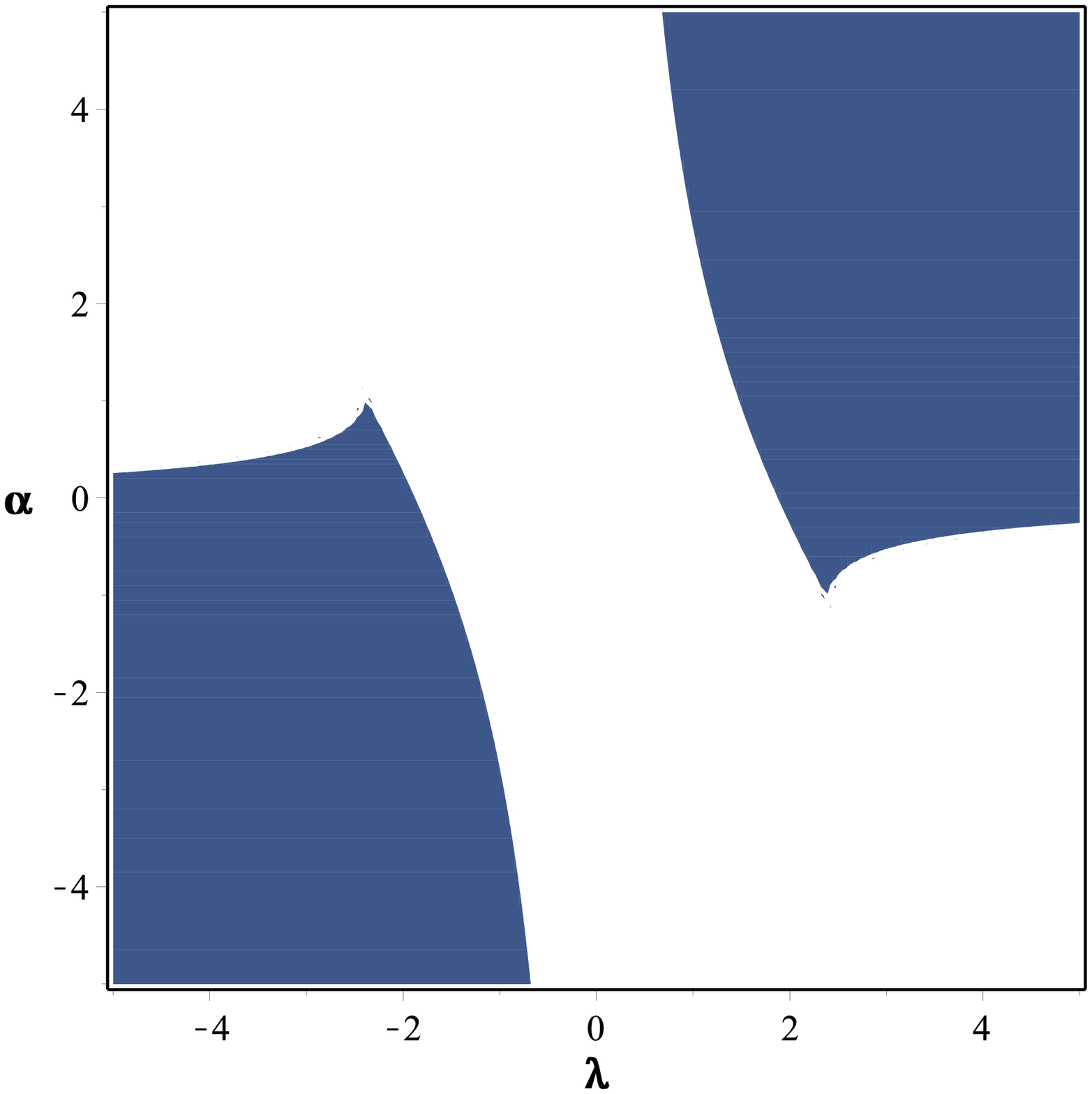}}
\caption{\label{fig8} Critical points $F_{\pm\mp}$ are stable nodes in the narrow
shaded regions of the $\lambda$-$\alpha$ plane (left panel), while
they are stable spirals in the shaded regions of the right panel.}
\end{figure*}

\begin{itemize}
\item \textbf{Critical point A:}\\
The critical point $A$ represents an attractor point for $\alpha<-\frac{\sqrt{6}}{2}$ and $\lambda>\sqrt{6}$. Otherwise it is saddle point in the phase space and a scalar field's kinetic energy term dominates the universe. In this case we have no late-time acceleration.

\item \textbf{Critical points $B_\pm$:}\\
Like as the fixed point $A$, the critical points $B_\pm$ show saddle points in the phase space. These points
belong to matter domination era. These points represent that the matter domination era is a transient phase with deceleration.
\item \textbf{Critical points $C_{\pm}$:}\\
The critical points $C_{\pm}$ show a solution with matter density term domination and a scalar field's kinetic energy term domination. As we see this contribution depends on the value of $\alpha$. But the behavior of two critical points depends on the value of  $\lambda$ and $\alpha$. If we consider $\lambda^{2}<\frac{9}{10}$ and $\lambda\alpha-\alpha^2>\frac{3}{2}$ these two critical points are attractors,
otherwise they will be saddle points. However, in both cases
we can not reach accelerated phase of expansion.
\item \textbf{Critical points $D_{\pm}$:}\\
The critical points $D_{\pm}$ denote either a solution with a potential
energy term domination or a scalar field's kinetic energy term domination. As
we see this contribution depends on the value of $\lambda$. These two critical points
behave like attractor points in the phase space if we consider $\lambda^{2}<6$ and $\lambda^2-\lambda\alpha<3$.
But if we consider $\lambda^2-\lambda\alpha>3$ these two fixed points will be saddle. However, in both cases
supposing $\lambda^{2}<2$ leads to reach accelerating phase of expansion.
\item \textbf{Critical points $E_\pm$:}\\
These critical points represent cosmological constant domination. Unfortunately in this case the eigenvalues are indefinite and one cannot understand the behavior of
the fixed points $E_\pm$.
\item \textbf{Critical points $F_{\pm\mp}$:}\\
The critical points $F_{\pm\mp}$ are scaling solutions with accelerated expansion and naturally the coincidence problem can be alleviate in this case. Depending on the values of the eigenvalues there are two situations: If we choose $\lambda$ and $\alpha$ parameters from the shaded regions in the left panel of figure \ref{fig8}, the critical points will be attractor nodes. But, considering $\lambda$ and $\alpha$ parameters from the shaded region in the right panel of figure \ref{fig8}, the critical points will be stable spiral. It means that the eigenvalues are complex numbers with negative real parts. For $\frac{\alpha}{\lambda+\alpha}>\frac{1}{3}$ there is an accelerated expansion phase. In fact, our analysis verifies that $\frac{\Omega_{m}}{\Omega_{\varphi}}<1$ and $\omega_{totc}<-\frac{1}{3}$. The phase portrait for this case is the same as the one illustrated in figure $1$. It shows that all the trajectories converge to the attractor points $F_{\pm\mp}$. By choosing correct values of quantities $\lambda$ and $\alpha$, we obtain the current value of dark matter density, $\Omega_{m}$, that is in agreement with the recent data from Planck2015 \cite{planck2015}, that is,  $\Omega_{m} = 0.3089\pm0.0062$
from TT, TE, EE+lowP+lensing+ext data.
 \end{itemize}

\begin{table*}
\begin{small}
\caption{\label{tab:5} Properties of the critical points for quintessence field. }
\begin{tabular}{cccccc}\\
\hline\hline \\
$(x_{2c},x_{3c},x_{4c})$ & Existence & Stability & $\Omega_\varphi$ & $\omega_{totc}$ & $\ddot{a}_c>0$\\\\
\hline\\ $A(0,0,0)$ & $\forall\lambda,~\alpha$ & attractor point if $\alpha<-\frac{\sqrt{6}}{2}$ and & 1 & 1 & No \\
             & & $\lambda>\sqrt{6}$; otherwise saddle point & & & \\\\
$B_\pm(0,\pm1,0)$ & $\forall\lambda,~\alpha$ & saddle point & 0 & 0 & No \\\\
&  & attractor point if $\alpha^2<\frac{9}{10}$  &  &  &  \\
     $C_\pm(0,\pm\sqrt{1-\frac{2\alpha^2}{3}},0)$ & $\forall\lambda$ and $\alpha^2\leq\frac{3}{2}$ & and $\lambda\alpha-\alpha^2>\frac{3}{2};$ & $\frac{2}{3}\alpha^2$ & $\frac{2}{3}\alpha^2$ & No \\
                       & & otherwise saddle point  &  &  &  \\\\
&  & attractor point if $\lambda^2<6$  &  &  &  \\
     $D_\pm(\pm\sqrt{1-\frac{\lambda^2}{6}},0,0)$ & $\forall\alpha,~\lambda^2\leq6$ & and $\lambda^2-\alpha\lambda<3$; & 1 & $-1+\frac{1}{3}\lambda^2$ & Yes if \\
                       & & saddle point if  $\lambda^2<6$  &  &  & $\lambda^2<2$ \\
                       & & and $\lambda^2-\alpha\lambda>3$ &  &  &  \\\\
$E_{\pm}(\pm1,0,x_4)$ & $\forall\lambda,~\alpha$ & undefined & 1 & -1 & Yes \\\\
$F_{\pm}(\pm\frac{\sqrt{\alpha^2+\alpha\lambda+\frac{3}{2}}}{\alpha+\lambda},$ & $0\leq\alpha^2+\alpha\lambda+\frac{3}{2}\leq(\alpha+\lambda)^2$ & attractor point (fig. \ref{fig8}, left panel) & $\frac{\alpha^2+\alpha\lambda+3}{(\alpha+\lambda)^2}$ & $-\frac{\alpha}{\alpha+\lambda}$ & Yes if \\
            $\pm\frac{\sqrt{\lambda^2+\alpha\lambda-3}}{\alpha+\lambda},0)$ & $0\leq\lambda^2+\alpha\lambda-3\leq(\alpha+\lambda)^2$ & attractor spiral (fig. \ref{fig8}, right panel) & & & $\frac{\alpha}{\lambda+\alpha}>\frac{1}{3}$ \\\\
$G_{\pm}(0,\pm\frac{\sqrt{-2\alpha^2-6}}{\alpha},$ & not exists  & - & - & - & - \\
               $ \pm\sqrt{2\alpha^2+3})$ &  &  &  &  &  \\\\
\hline \hline
\end{tabular}
\end{small}
\end{table*}

\begin{table*}
\begin{small}
\caption{\label{tab:6} The eigenvalues ($\theta_i$'s) of the critical points for quintessence field.}
\begin{tabular}{cc}\\
\hline\hline \\
point$(x_{2c},~x_{3c},~x_{4c})$ & $\theta_1$, $\theta_2$, $\theta_3$\\\\
\hline\\ $A(0,0,0)$ & $-3,~\frac{3}{2}+\frac{\sqrt{6}}{2}\alpha,~3-\frac{\sqrt{6}}{2}\lambda$ \\\\
 $B_\pm(0,\pm1,0)$ & $-\frac{3}{2}$,~$\frac{3}{2}$,~undefined \\\\
 $C_\pm(0,\pm\sqrt{1-\frac{2\alpha^2}{3}},0)$  & $-\alpha^2-\frac{3}{2},~\alpha^2-\lambda\alpha+\frac{3}{2},~5\alpha^2-\frac{9}{2}$ \\\\
 $D_\pm(\pm\sqrt{1-\frac{\lambda^2}{6}},0,0)$ & $\frac{1}{2}\lambda^2-3$, $\frac{\lambda^2-\alpha\lambda-3}{2}$, $-\frac{1}{2}\lambda^2$ \\\\
 $E_{\pm}(\pm1,0,x_4)$ & undefined \\\\
 $F_{\pm}(\pm\frac{\sqrt{\alpha^2+\alpha\lambda+\frac{3}{2}}}{\alpha+\lambda},\pm\frac{\sqrt{\lambda^2+\alpha\lambda-3}}{\alpha+\lambda},0)$ & $-\frac{3\lambda}{2(\alpha+\lambda)}$,$~\frac{-\frac{3}{2}\alpha-\frac{3}{4}\lambda\pm\frac{1}{4}\sqrt{-48\alpha^3\lambda-96\alpha^2\lambda^2
 -48\alpha\lambda^3+180\alpha^2+108\alpha\lambda-63\lambda^2+216}}{\alpha+\lambda}$\\\\
 $G_{\pm}(0,\pm\frac{\sqrt{-2\alpha^2-6}}{\alpha},\pm\sqrt{2\alpha^2+3}) $ & - \\\\
\hline \hline
\end{tabular}
\end{small}
\end{table*}

\subsection{The phase space with a phantom field}

Solving equations (\ref{x'2}), (\ref{x'4}) and (\ref{x'33}) with $\epsilon=-1$, we reach at seven critical points $(A, B, C, D, E, F, G)$ in our system. These critical points and stability around them are summarized in tables \ref{tab:7} and \ref{tab:8}. Now we investigate the properties of each critical point separately. In all calculations like previous subsection we suppose the condition $\gamma=1$.

\begin{itemize}
\item \textbf{Critical point A:}\\
The fixed point $A$ shows a saddle point in a scalar field's kinetic energy dominated universe and
in this case we have no late-time acceleration.
\item \textbf{Critical points $B_\pm$:}\\
Like the previous subsection, the critical points $B_\pm$ show saddle points in the matter dominated phase space and represent that
the matter domination era is a transient phase.
\item \textbf{Critical points $E_\pm$:}\\
These critical points represent cosmological constant domination era. Once again in this case the eigenvalues are indefinite and one can not understand the behavior of the fixed points $E_\pm$.
\item \textbf{Critical points $F_{\pm\mp}$:}\\
The critical points $F_{\pm\mp}$ show either a solution with matter density term plus a scalar field's kinetic energy term domination or just a potential term domination. But according to their eigenvalues, these points are saddle points and could not be attractor solutions for late time acceleration.
\item \textbf{Critical points $G_{\pm\mp}$:}\\
The critical points $G_{\pm\mp}$ show a solution with cosmological constant domination. These points also are saddle points and could not be considered as an attractor solution for late time acceleration.
\end{itemize}

\begin{table*}
\begin{small}
\caption{\label{tab:7} Properties of the critical points for phantom field. }
\begin{tabular}{cccccc}\\
\hline\hline \\
$(x_{2c},x_{3c},x_{4c})$ & Existence & Stability & $\Omega_\varphi$ & $\omega_{totc}$ & $\ddot{a}_c>0$\\\\
\hline\\ $A(0,0,0)$ & $\forall\lambda,~\alpha$ & saddle point & 1 & 1 & No \\\\
$B_\pm(0,\pm1,0)$ & $\forall\lambda,~\alpha$ & saddle point & 0 & 0 & No \\\\
$C_\pm(0,\pm\sqrt{1+\frac{2\alpha^2}{3}},0)$ & not exists & - & - & - \\\\
$D_\pm(\pm\sqrt{1+\frac{\lambda^2}{6}},0,0)$ & $\forall\lambda,~\alpha$ & saddle point & 1 & $-1-\frac{1}{3}\lambda^2$ & Yes \\\\
$E_{\pm}(\pm1,0,x_4)$ & $\forall\lambda,~\alpha$ & undefined & 1 & -1 & Yes \\\\
$F_{\pm}(\pm\frac{\sqrt{\alpha^2+\alpha\lambda-\frac{3}{2}}}{\alpha+\lambda},\pm\frac{\sqrt{\lambda^2+\alpha\lambda+3}}{\alpha+\lambda},0)$ & $\alpha^2+\alpha\lambda-\frac{3}{2}\geq0$ & saddle point & $\frac{\alpha^2+\alpha\lambda-3}{(\alpha+\lambda)^2}$ & $-\frac{\alpha}{\alpha+\lambda}$ & Yes if \\
   & $0\leq\lambda^2+\alpha\lambda+3\leq(\alpha+\lambda)^2$ &  & & & $\frac{\alpha}{\lambda+\alpha}>\frac{1}{3}$\\\\
$G_{\pm}(0,\pm\frac{\sqrt{-2\alpha^2+6}}{\alpha},\pm\sqrt{2\alpha^2-3})$ & $2\leq\alpha^2\leq3$ & saddle point & $\frac{3(\alpha^2-2)}{\alpha^2}$ & -1 & Yes \\\\
\hline \hline
\end{tabular}
\end{small}
\end{table*}

\begin{table*}
\begin{small}
\caption{\label{tab:8} The eigenvalues ($\theta_i$'s) of the critical points for phantom field.}
\begin{tabular}{cc}\\
\hline\hline \\
point$(x_{2c},~x_{3c},~x_{4c})$ & $\theta_1$, $\theta_2$, $\theta_3$\\\\
\hline\\
$A(0,0,0)$ & $-3,~\frac{3}{2}+\frac{\sqrt{-6}}{2}\alpha,~3-\frac{\sqrt{-6}}{2}\lambda$ \\\\
$B_\pm(0,\pm1,0)$ & $-\frac{3}{2}$,~$\frac{3}{2}$,~undefined \\\\
$C_\pm(0,\pm\sqrt{1+\frac{2\alpha^2}{3}},0)$  & - \\\\
$D_\pm(\pm\sqrt{1+\frac{\lambda^2}{6}},0,0)$ & $\frac{1}{2}\lambda^2$, $-\frac{5}{2}\lambda^2-9$, $-\frac{1}{2}\lambda^2+\frac{1}{2}\alpha\lambda-\frac{3}{2}$ \\\\
$E_{\pm}(\pm1,0,x_4)$ & undefined \\\\
$F_{\pm}(\pm\frac{\sqrt{\alpha^2+\alpha\lambda-\frac{3}{2}}}{\alpha+\lambda},\pm\frac{\sqrt{\lambda^2+\alpha\lambda+3}}{\alpha+\lambda},0)$ & $-\frac{3\lambda}{2(\alpha+\lambda)}$,$~\frac{-\frac{3}{2}\alpha-\frac{3}{4}\lambda\pm\frac{1}{4}\sqrt{48\alpha^3\lambda+96\alpha^2\lambda^2
    +48\alpha\lambda^3+180\alpha^2+108\alpha\lambda-63\lambda^2-216}}{\alpha+\lambda}$\\\\
$G_{\pm}(0,\pm\frac{\sqrt{-2\alpha^2+6}}{\alpha}, \pm\sqrt{2\alpha^2-3})$ & $-\frac{3\lambda}{2\alpha},~\mp\frac{3}{2}\frac{\pm8\alpha^4\mp21\alpha^2+
   \sqrt{120\alpha^6-531\alpha^4+702\alpha^2-243}\pm9}{8\alpha^4-21\alpha^2+9}$\\\\
\hline \hline
\end{tabular}
\end{small}
\end{table*}

\section{Perturbations in quintessence}

Now we study cosmological perturbations in this setup with a quintessence field that is coupled
non-minimally with dark matter and also its derivatives are coupled to the background
curvature. In this case the energy transfers from dark energy to dark matter. We investigate the analytical solution of matter perturbations in this coupled
scenario and compare the results with matter perturbation solutions without interaction
between the dark sectors. Equations (\ref{pe1}), (\ref{pe2}), (\ref{pe3}) and (\ref{pe4}) are valid in this situation too, but equations (\ref{deltam}) and (\ref{vdot}) now read as follows
\begin{eqnarray}\label{deltadeltam}
&&\delta(F(\varphi)\rho_{m}\dot{)}+3H\delta\Big(F(\varphi)(\rho_{m}+p_{m})\Big)=\nonumber\\
&&~~~~F(\varphi)(\rho_{m}+p_{m})\Big(-3\dot{\psi}+\frac{\nabla^2}{a}\nu_{m}\Big)\nonumber\\
&&-\delta F'(\varphi)\rho_{m}\dot{\varphi}-F'(\varphi)\delta\rho_{m}\dot{\varphi}-F'(\varphi)\rho_{m}\delta\dot{\varphi},
\end{eqnarray}

\begin{eqnarray}
\dot{\nu}_{m}+\Big[(1-3\omega_{m})H+\frac{F'(\varphi)}{F(\varphi)}\dot{\varphi}\Big]\nu_{m}=\nonumber\\
\frac{1}{a}\Big[A+\frac{\omega_m}{1+\omega_{m}}\delta_m
-\frac{F'(\varphi)}{F(\varphi)}\frac{\delta\varphi}{1+\omega_m}\Big].
\end{eqnarray}
where a prime denotes derivative with respect to $\varphi$.
We investigate the evolution of perturbations on sub-Hubble scales. We calculate $\delta_m\equiv\delta(F(\varphi)\rho_m)/F(\varphi)\rho_m$ in order to derive matter perturbations. In which follows we consider $\omega_m$ as a constant and using the Fourier-transformed equations (\ref{Fourier1}-\ref{Fourier3}) and Bardeens potentials (\ref{Bardeen1}) and (\ref{Bardeen2}), we find

\begin{eqnarray}\label{deltadeltamm}
\dot{\delta}_{m}+2\frac{F'(\varphi)}{F(\varphi)}\dot{\varphi}\delta_m=(1+\omega_{m})(3\dot{\Psi}-\frac{k^2}{a}\nu_{m})\nonumber\\
-\frac{\delta F'(\varphi)}{F(\varphi)}\dot{\varphi}-\frac{F'(\varphi)}{F(\varphi)}\delta\dot{\varphi},
\end{eqnarray}

\begin{eqnarray}
\dot{\nu}_{m}+\Big[H(1-3\omega_{m})+\frac{F'(\varphi)}{F(\varphi)}\dot{\varphi}\Big]\nu_{m}=\nonumber\\
\frac{1}{a}\Big[\Phi+\frac{\omega_{m}}{1+\omega_{m}}\delta_m
-\frac{F'(\varphi)}{F(\varphi)}\frac{\delta\varphi}{1+\omega_m}\Big]\label{v}.
\end{eqnarray}
Taking derivative from Eq. (\ref{deltadeltamm}) and eliminating $\nu_m$ from the two above equations, we find

\begin{eqnarray}\label{deltadeltammm}
&&\ddot{\delta}_{m}+\bigg[(2-3\omega_m)H+3C\bigg]\dot{\delta}_m+\Bigg[\frac{k^2}{a^2}\omega_m+2\dot{C}+\nonumber\\
&&2\Big[(2-3\omega_m)H+C\Big]C\Bigg]\delta_m+\frac{k^2}{a^2}\bigg[\Phi-\frac{E}{1+\omega_m}\bigg](1+\omega_m)= \nonumber\\
&&3(1+\omega_m)\Bigg[\ddot{\Psi}+\Big[(2-3\omega_m)H+C\Big]
\Big[\dot{\Psi}-\frac{D+B}{3(1+\omega_m)}\Big]\Bigg]\nonumber\\
&&~~~~~~~~~~~~~~~~~~~~~~~~~~~~~-\dot{D}-\dot{B}.
\end{eqnarray}
where

\begin{eqnarray}
D=\frac{\delta F'(\varphi)}{F(\varphi)}\dot{\varphi},\,\,\,\,\,\,B=\frac{F'(\varphi)}{F(\varphi)}\delta\dot{\varphi},\nonumber\\
C=\frac{F'(\varphi)}{F(\varphi)}\dot{\varphi},\,\,\,\,\,\,E=\frac{F'(\varphi)}{F(\varphi)}\delta\varphi,\nonumber
\end{eqnarray}

\begin{eqnarray}
&&\dot{D}=\frac{F'''(\varphi)\delta\varphi}{F(\varphi)}\dot{\varphi}^2+\frac{F''(\varphi)}{F(\varphi)}\delta\dot{\varphi}\dot{\varphi}
+\frac{F''(\varphi)\delta\varphi}{F(\varphi)}\ddot{\varphi}\nonumber\\
&&~~~~~~~~~~~~~~~-\frac{F''(\varphi)F'(\varphi)\delta\varphi\dot{\varphi}^2}{F^{2}(\varphi)},\nonumber
\end{eqnarray}

\begin{eqnarray}
\dot{B}=\frac{F''(\varphi)}{F(\varphi)}\delta\dot{\varphi}\dot{\varphi}
+\frac{F'(\varphi)}{F(\varphi)}\delta\ddot{\varphi}-\frac{F'^2(\varphi)\dot{\varphi}}{F^2(\varphi)}\delta\dot{\varphi},\nonumber
\end{eqnarray}

\begin{eqnarray}
\dot{C}=\frac{F''(\varphi)}{F(\varphi)}\dot{\varphi}^2+\frac{F'(\varphi)}{F(\varphi)}\ddot{\varphi}
-\frac{F'^2(\varphi)\dot{\varphi}}{F^2(\varphi)}\dot{\varphi}^2.\nonumber
\end{eqnarray}

We investigate non-relativistic matter $(\omega_m=0)$ on scales which are much smaller than the Hubble radius $(k\gg aH)$. In this case, Eq. (\ref{deltadeltammm}) takes the following form

\begin{eqnarray}
&&\ddot{\delta}_{m}+\bigg[2H+3\frac{F'(\varphi)}{F(\varphi)}\dot{\varphi}\bigg]\dot{\delta}_m+
2\Bigg[\frac{F''(\varphi)}{F(\varphi)}\dot{\varphi}^2+\frac{F'(\varphi)}{F(\varphi)}\ddot{\varphi}\nonumber\\
&&-\frac{F'^2(\varphi)\dot{\varphi}}{F^2(\varphi)}\dot{\varphi}^2+\Big[2H+
\frac{F'(\varphi)}{F(\varphi)}\dot{\varphi}\Big]\frac{F'(\varphi)}{F(\varphi)}\dot{\varphi}\Bigg]\delta_m\nonumber\\
&&~~~~~~~~~~+\frac{k^2}{a^2}\bigg[\Phi-\frac{F'(\varphi)}{F(\varphi)}\delta\varphi\bigg]=0,
\end{eqnarray}
Considering $F'(\varphi)=\alpha F(\varphi)$, $\dot{\varphi}$ in terms of the dimensionless parameter from (\ref{di}) and $\delta\varphi$ from perturbed Klein-Gordon equation, we get

\begin{eqnarray}
&&\ddot{\delta}_{m}+\Big(2H+3\sqrt{6}\alpha Hx_1\Big)\dot{\delta}_m-
\frac{3}{2}H^2\Omega_m\Big(1+2\alpha^2\nonumber\\
&&+2\alpha\ddot{\varphi}+4\sqrt{6}\alpha H^2x_1+12\alpha^2H^2x_1^2\Big)\delta_m=0.
\end{eqnarray}
Using the relation $d/dN=(1/H)(d/dt)$ we obtain

\begin{eqnarray}\label{variationn}
&&\frac{d^2\delta_m}{dN^2}+\bigg[2+\frac{1}{H}\frac{dH}{dN}+3\sqrt{6}\alpha x_1\bigg]\frac{d\delta_m}{dN}
+\bigg[-\frac{3}{2}\Omega_m\big(\nonumber\\
&&1+2\alpha^2\big)+2\alpha\frac{\ddot{\varphi}}{H^2}+4\sqrt{6}\alpha x_1+12\alpha^2x_1^2\bigg]\delta_m=0.
\end{eqnarray}

\subsection{Perturbations in the matter domination era}

\begin{figure*}
\flushleft\leftskip0em{
\includegraphics[width=.45\textwidth,origin=c,angle=0]{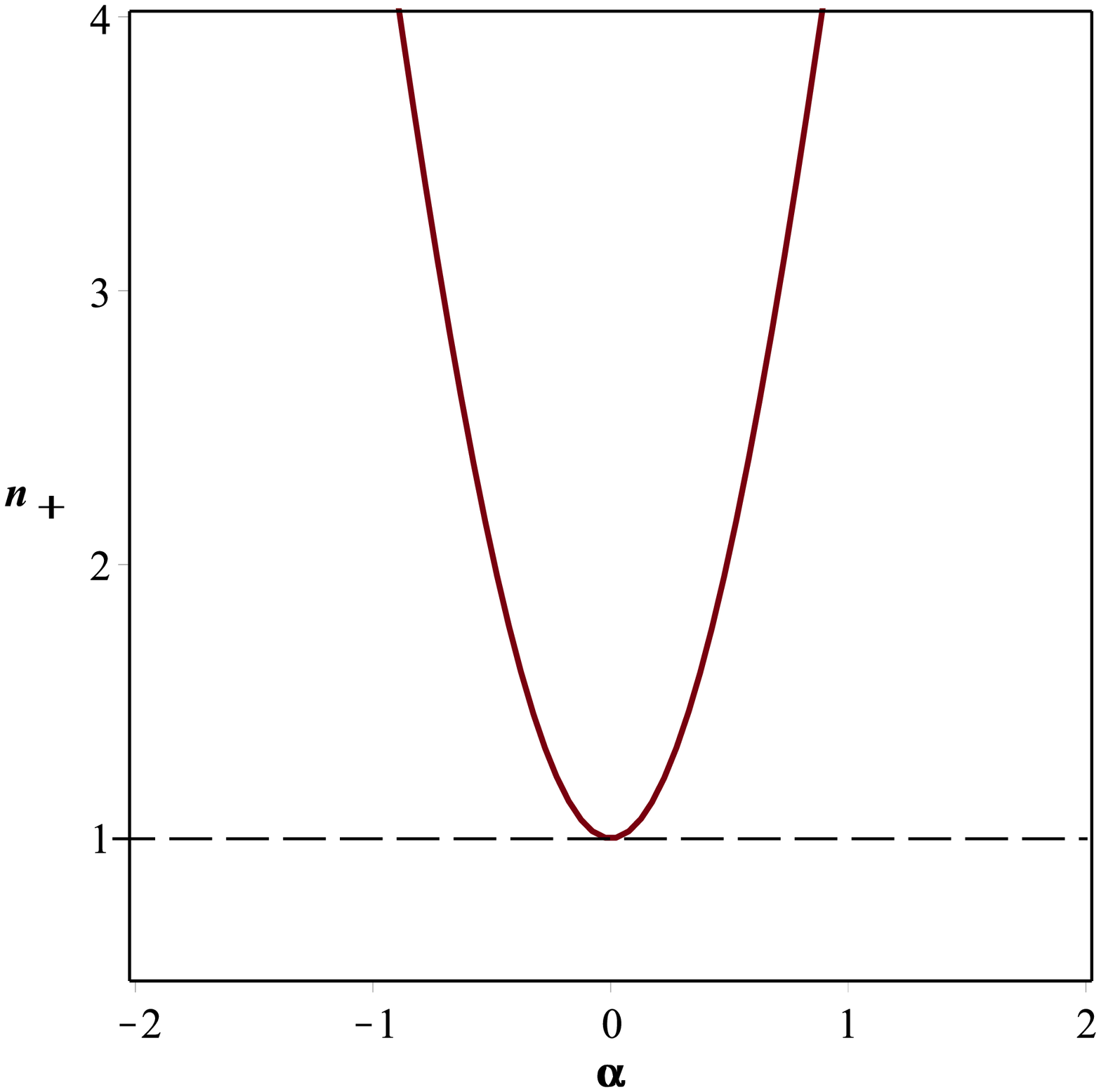}
\hspace{0.5cm}
\includegraphics[width=.45\textwidth,origin=c,angle=0]{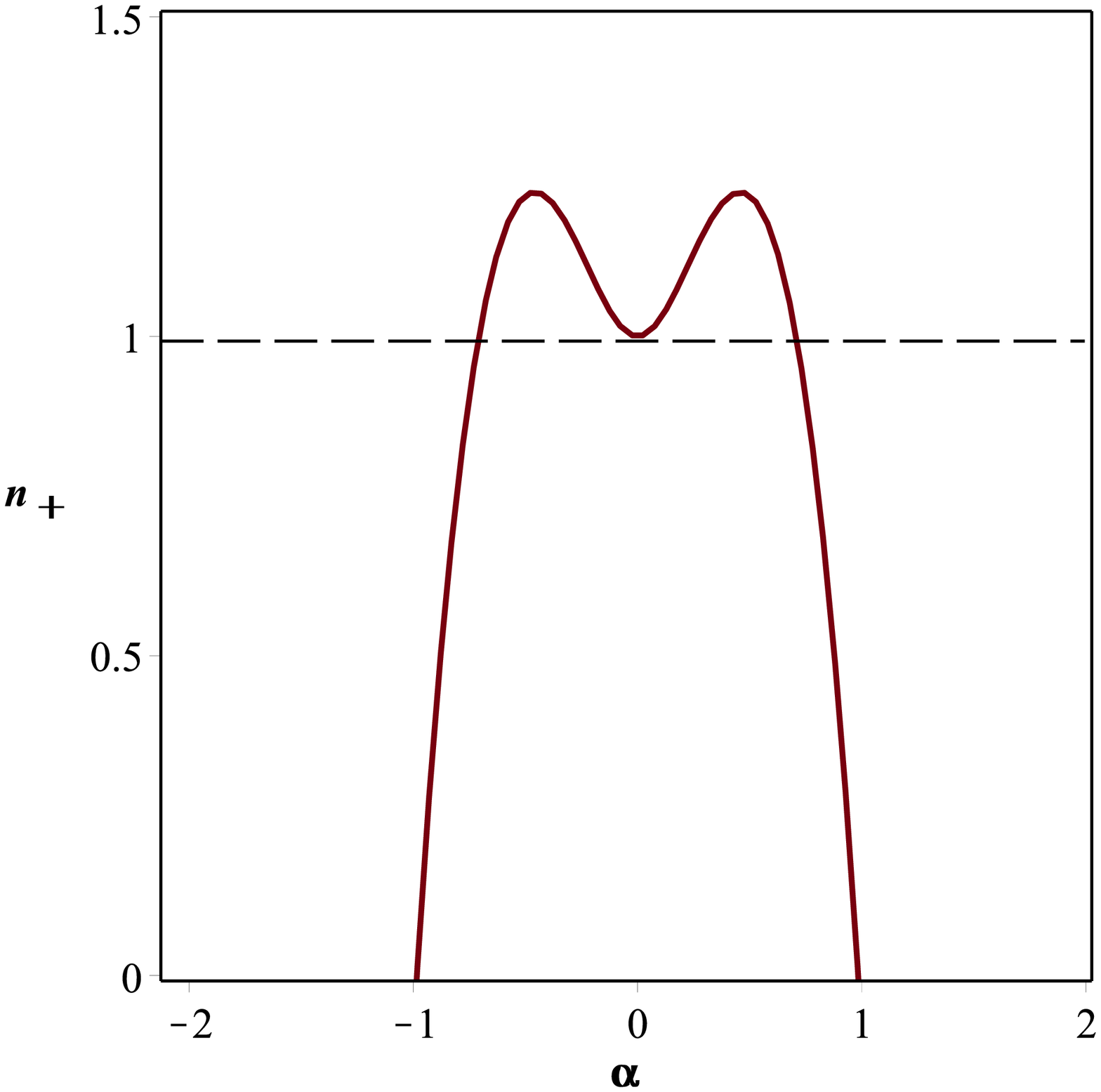}}
\caption{\label{fig9} The growth rate of matter perturbations versus $\alpha$ in the matter domination era. The left panel is drawn for $x_1=-\sqrt{\frac{2}{3}}\alpha$ and the right panel for $x_1=+\sqrt{\frac{2}{3}}\alpha$.}
\end{figure*}

In our model by choosing transient regime according to the critical points $C$ in table \ref{tab:5}, (that is $\alpha^2<\frac{3}{2}$ and $-\alpha^2+\lambda\alpha<\frac{3}{2}$), one has

\begin{eqnarray}
&&x_1=\pm\sqrt{\frac{2}{3}}\alpha,~~~,x_3=\pm\sqrt{1-\frac{2}{3}\alpha^2},~~~\Omega_\varphi=\frac{2}{3}\alpha^2\nonumber\\
&&~~~\omega_{totc}=\frac{2}{3}\alpha^2,~~~\frac{\ddot{\varphi}}{H^2}=-3\alpha x_3^2-3\sqrt{6}x_1
\end{eqnarray}
and using the relation

\begin{equation}\label{dh/dnn}
\frac{1}{H}\frac{dH}{dN}=-\frac{3}{2}(1+\omega_{tot}),
\end{equation}
and $\Omega_m=1-\Omega_{\varphi}$, the equation (\ref{variationn}) takes the following forms

\begin{eqnarray}\label{x-}
&&\frac{d^2\delta_m}{dN^2}+\Big(\frac{1}{2}-7\alpha^2\Big)\frac{d\delta_m}{dN}+\Big(14\alpha^4-4\alpha^2-\frac{3}{2}\Big)\delta_m=0\nonumber\\ &&~~~~~~~~~~~~~~~~~~~for\,\,x_1=-\sqrt{\frac{2}{3}}\alpha
\end{eqnarray}

\begin{eqnarray}\label{x+}
&&\frac{d^2\delta_m}{dN^2}+\Big(\frac{1}{2}+5\alpha^2\Big)\frac{d\delta_m}{dN}+\Big(-\frac{3}{2}+6\alpha^4\Big)\delta_m=0\nonumber\\
&&~~~~~~~~~~~~~~~~~~~for\,\,x_1=+\sqrt{\frac{2}{3}}\alpha
\end{eqnarray}
Solving the above equation(s) we reach the general solution of the form

\begin{equation}\label{anaal}
\delta_m=C_1a^{n_+}+C_2a^{n_-}.
\end{equation}
For equation (\ref{x-}) we have
\begin{equation}
n_\pm=\frac{7}{2}\alpha^2-\frac{1}{4}\pm\frac{1}{4}\sqrt{-28\alpha^4+36\alpha^2+25}
\end{equation}
We just consider the $n_+$ case. It is easy to see that in the minimal case, when $\alpha=0$ then $\delta_{m}\propto a$ in matter domination era similar to the case mentioned in Ref. \cite{Copeland}. But, if we consider interaction between dark energy and dark matter, the growth rate is higher than the minimal case. This is shown in the left panel of figure \ref{fig9}.

For equation (\ref{x+}) we have

\begin{equation}
n_\pm=-\frac{5}{2}\alpha^2-\frac{1}{4}\pm\frac{1}{4}\sqrt{-124\alpha^4+212\alpha^2+25}
\end{equation}
Once again we just consider the $n_+$ case. With the same reason, when $\alpha=0$ then $n_+=1$ and $\delta_{m}\propto a$. But, if we consider interaction between dark energy and dark matter, the growth rate is lower than the minimal case and this depends on the values of $\alpha$. This is shown in the right panel of figure $10$. For small values of $\alpha$ it is slightly more than $1$, but for other values of $\alpha$, the growth rate of matter perturbation is less than the minimal case. We note that in the matter domination era the values of $\alpha$ are extremely small. Therefore, the growth rate of matter perturbation is slightly more than the minimal case.

\subsection{Perturbations in the scaling solution domination era}

Now we investigate the growth rate of matter perturbations in the scaling solution era. From table \ref{tab:5}, for the critical points $F_{\pm\mp}$ that have scaling solutions, we have

\begin{eqnarray}
&&~~~~~~~~~~~~~~~x_1=\pm\frac{1}{2}\frac{\sqrt{6}}{\alpha+\lambda},\nonumber\\
&&\Omega_{\varphi}=\frac{\alpha^2+\alpha\lambda+3}{(\alpha+\lambda)^2},~~~\omega_{totc}=\frac{-\alpha}{\alpha+\lambda}
\end{eqnarray}

By replacing the above relations, using $\Omega_m=1-\Omega_{\varphi}$ and equation (\ref{dh/dnn}) in equation (\ref{variationn}), we get

\begin{eqnarray}\label{sdeltampp}
&&\frac{d^2\delta_m}{dN^2}+\bigg[\frac{1}{2}+\frac{21}{2}(\frac{\alpha}{\alpha+\lambda})\bigg]\frac{d\delta_m}{dN}
+\Bigg[-\frac{3}{2}\bigg(\frac{\alpha\lambda-\lambda^2-3}{(\alpha+\lambda)^2}\bigg)\nonumber\\
&&~~~~~~~~~~\Big(1+2\alpha^2\Big)+\frac{3\alpha(\lambda+10\alpha)}{(\alpha+\lambda)^2}\Bigg]\delta_m=0.\nonumber\\
&&~~~~~~~~~~~~~~~~~~~For\,\,\,\,  x_1=\frac{1}{2}\frac{\sqrt{6}}{\alpha-\lambda}
\end{eqnarray}

\begin{eqnarray}\label{sdeltamnn}
&&\frac{d^2\delta_m}{dN^2}+\bigg[\frac{1}{2}-\frac{15}{2}(\frac{\alpha}{\alpha+\lambda})\bigg]\frac{d\delta_m}{dN}
+\Bigg[-\frac{3}{2}\bigg(\frac{\alpha\lambda+\lambda^2-3}{(\alpha+\lambda)^2}\bigg)\nonumber\\
&&~~~~~~~~~~\Big(1+2\alpha^2\Big)+\frac{3\alpha(5\lambda+14\alpha)}{(\alpha+\lambda)^2}\Bigg]\delta_m=0.\nonumber\\
&&~~~~~~~~~~~~~~~~~~~For\,\,\,\,  x_1=-\frac{1}{2}\frac{\sqrt{6}}{\alpha-\lambda}
\end{eqnarray}

\begin{figure*}
\flushleft\leftskip0em{
\includegraphics[width=.45\textwidth,origin=c,angle=0]{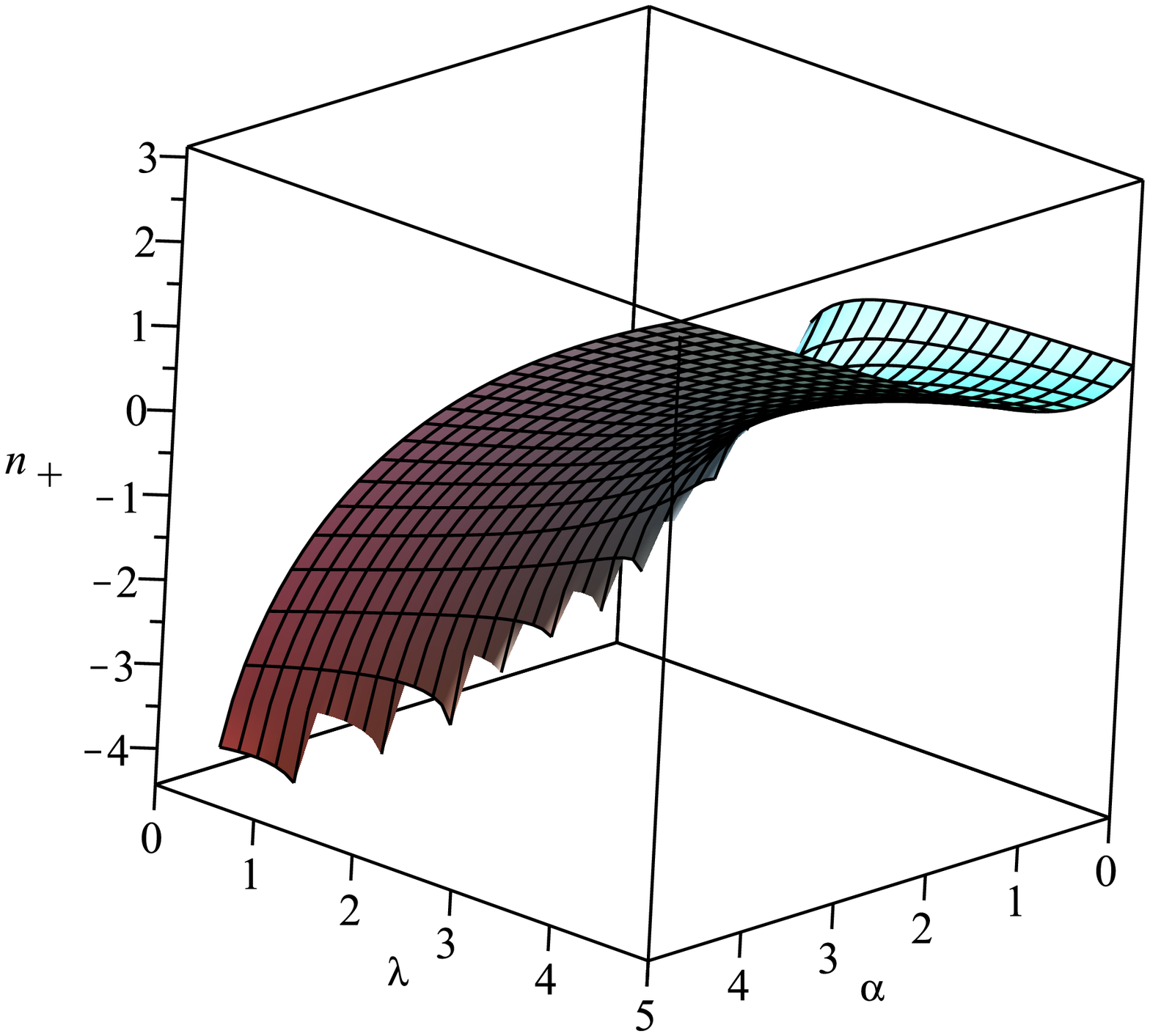}
\hspace{0.5cm}
\includegraphics[width=.45\textwidth,origin=c,angle=0]{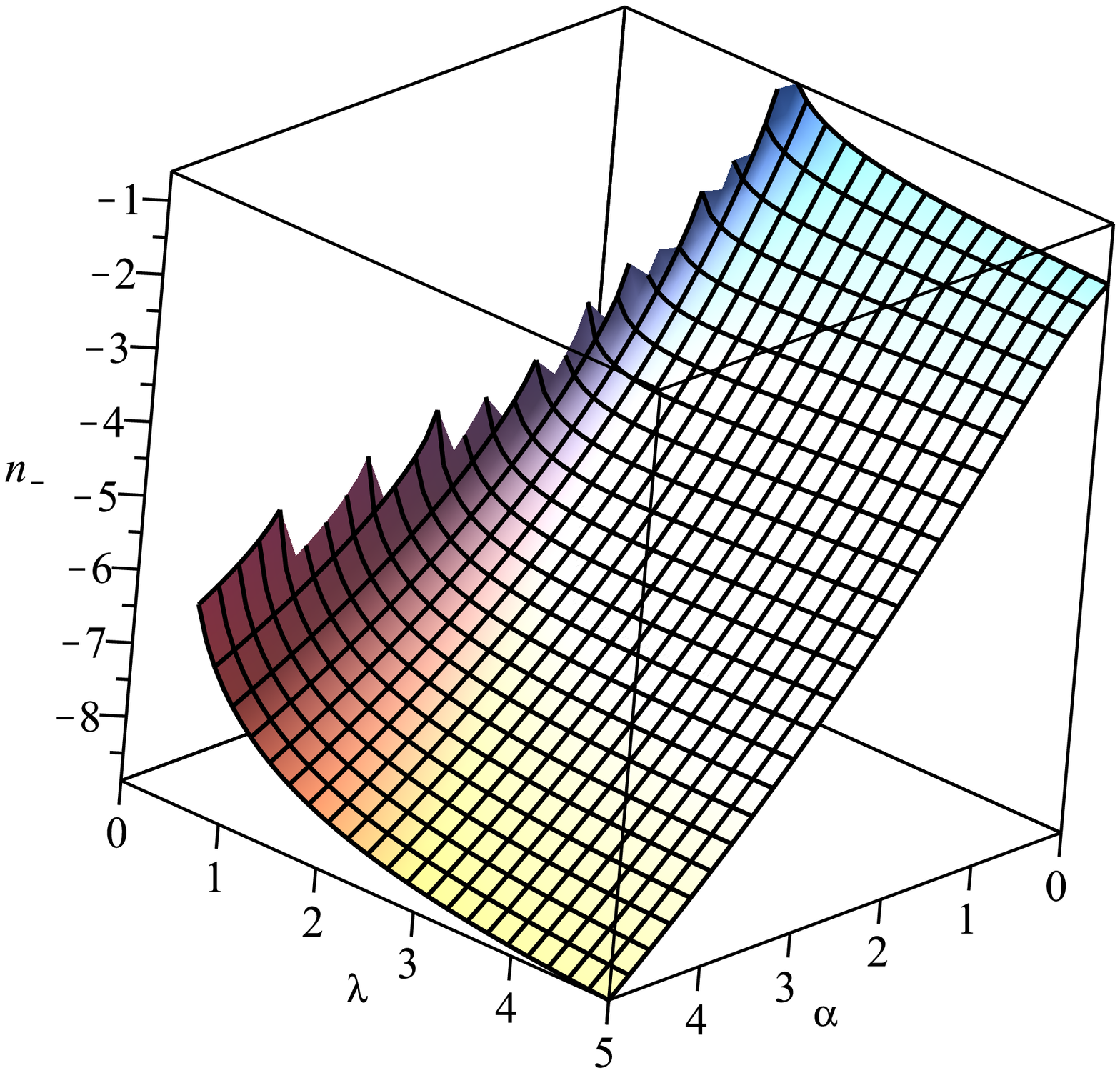}}
\caption{\label{fig10} The growth rate of matter perturbations versus $\alpha$ and $\lambda$ in the late-time scaling solution domination era.}
\end{figure*}

Solving the equation (\ref{sdeltampp}), we obtain the analytic solution like equation (\ref{anaal}) with the following $n_{\pm}$

\begin{eqnarray}
&&n_{\pm}=\frac{1}{4}\Bigg(\frac{-22\alpha-\lambda}{\alpha+\lambda}\nonumber\\
&&\pm\frac{\sqrt{48\alpha^3\lambda+48\alpha^2\lambda^2-140\alpha^2+20\alpha\lambda+25\lambda^2-72}}{\alpha+\lambda}\Bigg)\nonumber\\
\end{eqnarray}
Also by solving the equation (\ref{sdeltamnn}) we obtain the analytic solution like equation (\ref{anaal}) with the following $n_{\pm}$

\begin{eqnarray}
&&n_{\pm}=\frac{1}{4}\Bigg(\frac{14\alpha-\lambda}{\alpha+\lambda}\nonumber\\
&&\pm\frac{\sqrt{48\alpha^3\lambda+48\alpha^2\lambda^2-620\alpha^2-244\alpha\lambda+25\lambda^2-72}}{\alpha+\lambda}\Bigg)\nonumber\\
\end{eqnarray}
The above equations show that the growth rate of matter perturbations depends on the two quantities $\alpha$ and $\lambda$. As figure \ref{fig10} shows, for positive $\lambda$ and $\alpha$ (that we have derived in the previous section), we have two possible cases: $n_+>0$ and $n_-<0$. So, we consider just the $n_+$ case. For uncoupled dark energy and dark matter, $n_+\leq1$ (see \cite{Copeland}), but in our interacting model $n_+$ can become even larger than $1$ in both cases. This shows that in the interacting scenario the growth rate of perturbations is larger than the minimal non-interacting case.

\section{Conclusion}

In this paper we have studied cosmological dynamics of an extended gravitational theory that gravity is coupled
non-minimally with derivatives of a dark energy component and there is also an explicit and phenomenological
interaction between the dark energy and dark matter. This is a simple interacting dark energy model that has the potential to alleviate the coincidence problem.
In the first step we considered the direction of energy flow
from dark matter to dark energy. In this case for a quintessence field we have shown that there are critical points $F_{\pm\mp}$ that are attractor scaling solutions in phase space of the model. With this scaling solution the issue of cosmological coincidence can be alleviated. This happens by choosing $\lambda$ and $\alpha$ parameters from the shaded region in the left panel of figure \ref{fig1}. We obtained also the current value of the dark matter density parameter, $\Omega_{m}$, that is in agreement with the recent data from Planck2015 \cite{planck2015}, that is, $\Omega_{m} = 0.3089\pm0.0062$ from TT, TE, EE+lowP+lensing+ext data. Furthermore, these points represent that for positive $\lambda$ we have to consider negative $\alpha$ and coupling term behaves like a potential function. For a phantom field, the critical points $F_{\pm\mp}$ show that if we choose $\alpha$ and $\lambda$ from the shaded region in the left panel of figure \ref{fig2}, the critical points will be attractor nodes. We were able to obtain the current value of the dark matter density, $\Omega_{m}$, that is in agreement with the recent data from Planck2015 \cite{planck2015} by choosing $\alpha=-3.5$ and $\lambda=0.5$. Furthermore, the equation of state parameter of the dark energy gets a value which is close to the equation of state parameter of the dark energy from TT, TE, EE+lowP+lensing+ext data in Ref. \cite{planck2015} as $\omega=-1.019_{-0.080}^{+0.075}$.

The most important issue about points $F_{\pm\mp}$ is that the existence of the non-minimal derivative coupling provides the possibility of having attractor solution (or scaling solution) for the present universe in contrast with the previous works (such as \cite{Copeland}) that show jut the existence of future attractors. The existence of these scaling solutions sheds light on the issue of coincidence problem at this epoch. We note that in our case, the coincidence problem reduces to a simple choice of parameters in order to match dark energy/dark matter to recent observations. Since the accelerated scaling solution is an attractor in our case, no fine-tuning of initial conditions is needed. Nevertheless, the dynamics that produces such a scaling in the dark sectors possibly has other undesirable consequences (see \cite{Boehmer}). If we consider non-minimal derivative coupling without interaction between dark sectors, there is no attractor points \cite{Huang}. The critical points $G_{\pm\mp}$, for a narrow range of $\alpha$, show a solution with matter density and a scalar field's kinetic energy term domination. These points also, are critical points that carry some information about the role of the non-minimal derivative coupling in this setup. These solutions are attractor if we choose $\alpha$ and $\lambda$ from the shaded region in the right panel of figure \ref{fig2}. As we see from the right panel of figure \ref{fig2}, $\alpha$ and $\lambda$ parameters have the same signs in contrast with the critical points $F_{\pm\mp}$. To find a value of the dark matter density parameter, $\Omega_{m}$, that is in agreement with the recent data from Planck 2015 \cite{planck2015} with $\Omega_{m} = 0.3089\pm0.0062$, we have to consider $\alpha=1.61$ with any positive $\lambda$. Furthermore, the equation of state parameter of the dark energy reaches to $-1.85$, which is close to the best fit $\omega=-1.94$ for Planck+WMAP and to the best fit $\omega=-1.94$ for Planck+WMAP+high L \cite{planck2013}, $\omega=-1.54^{+0.62}_{-0.50}$ for TT and $\omega=-1.55^{+0.58}_{-0.48}$ for TE+EE in \cite{planck2015}. Our analysis has shown that in the absence of the non-minimal derivative coupling, there is no such a good agreement with data in our setup. In fact, existence of a non-minimal coupling between the derivatives of the dark energy component with curvature provides a better fit with observations in this setup. To proceed further, we have analyzed the classical stability of the solutions in the $\omega_{\varphi}'-\omega_{\varphi}$ phase-plane of the scalar fields with non-minimal derivative coupling for both quintessence and phantom field in two directions of energy flow between the dark sectors. The stability of solutions in $\omega_{\varphi}'-\omega_{\varphi}$ phase plane is independent on the direction of energy flow. The sound speed of the scalar field in order to get ride off the future big-rip singularity should be positive, that is, $c_{a}^{2}>0$. With this fact, the right panel of figure \ref{fig3} shows the regions of stability of classical solutions in $\omega_{\varphi}'-\omega_{\varphi}$ phase-plane. The region $I$ belongs to a quintessence phase, while region $III$ is for a phantom phase. These are the regions that the classical solutions are stable. Then we have studied the statefinder diagnostic in this non-minimal interacting model. By using statefinder diagnostic tool with $\{q, r\}$ and $\{q, s\}$ phase diagrams to distinguish between alternative dark energy models, we were able to see possible realization of the concordance $\Lambda$CDM phase in our setup. First we have considered a quintessence field where the right panel of figure \ref{fig4} illustrates the trajectories of $\{q, r\}$ and $\{q, s\}$ phase plane and $\omega_{\varphi}$ for the critical points $F_{\pm\mp}$ with two different values of $\lambda$ and $\alpha$ parameters in this case. The figures indicate that with different values of $\lambda$ and $\alpha$ parameters, the trajectories can evolve differently. Just by considering larger values of parameter $\alpha$, the trajectories of parameters will approach the $\Lambda$CDM model. So, the $\Lambda$CDM  is not in the cosmic history of quintessence field with non-minimal derivative coupling, at least for small values of the parameter $\alpha$. For phantom field the right panel of figure \ref{fig5} illustrates the trajectories of $\{q, r\}$ and $\{q, s\}$ phase plane and $\omega_{\varphi}$ for critical points $F_{\pm\mp}$ with two different values of the $\lambda$ and $\alpha$ parameters. Once again, these figures indicate that with different values of $\lambda$ and $\alpha$ parameters, the trajectories can evolve
differently. Considering smaller values of $\alpha$, the trajectories of parameters will approach the $\Lambda$CDM model and therefore $\Lambda$CDM belongs to the cosmic history with a phantom field with non-minimal derivative coupling.

As an important issue in cosmological dynamics we have studied the cosmological perturbations in this non-minimal interacting model with details and analytical solutions. In the first step we have studied the cosmological perturbations in our setup with a quintessence field that is coupled non-minimally with dark matter and also its derivatives are coupled to the background curvature. We investigated the analytical solution of matter perturbations in this coupled scenario and compared the results with matter perturbation solutions without interaction between the dark sectors in two cases, that is, matter domination and scaling solution era. We considered Bardeen's or gauge invariant potentials and non-relativistic matter $(\omega_m=0)$ on scales much smaller than the Hubble radius $(k\gg aH)$. In the matter domination era by considering interaction between dark energy and dark matter, for small values of $\alpha$ (note that in the matter domination era the values of $\alpha$ are extremely small), the growth rate of matter perturbation is slightly more than the value for the minimal case (see for instance Ref. \cite{Copeland} where $\delta_{m}\propto a$). This feature is shown in the right panel of figure \ref{fig6}. For growth rate of matter perturbations in the scaling solution era, (as is shown in table \ref{tab:1}, for the critical points $F_{\pm}$ there are scaling solutions), the growth rate of matter perturbations depends on two quantities $\alpha$ and $\lambda$. As we have shown in the right panel of figure \ref{fig7}, for positive $\lambda$ and negative $\alpha$, $n_{+}>0$ and $n_{-}<0$. By considering just $n_{+}$ case, for uncoupled dark energy and dark matter $n_+\leq1$ (see for instance \cite{Copeland}), but in our interacting model $n_{+}$ becomes larger than $1$ in both cases. This shows that in the interacting scenario the growth rate of perturbations is larger than the minimal non-interacting case.

We have extended our analysis for the case that the direction of the energy flow gets reversed. That is, the energy flows from dark energy to dark matter. The totality of the analysis is the same, but there is a significant difference that in contrast to the previous case, now both $\lambda$ and $\alpha$ parameters have the same positive signs for quintessence field. That is to say, potential and the field do not grow in the same manner. For the phantom field in this case, in contrast to the previous case, there is no scaling solution. For perturbations, the growth rate has no considerable difference from the case with opposite direction of the energy flow. Finally, we note that near coincidence of dark sectors energy densities in our case does no longer depend on the initial conditions. It depends only on the interaction between the dark sectors and coupling constants. Nevertheless, while the phase-space trajectory the universe follows from some point onward is unique and independent of the initial conditions, the current position of universe on this trajectory depends on them as usually happens in all cosmological models. In summary, an explicit interaction between the dark sectors in the presence of non-minimal derivative coupling between dark energy and curvature realizes scaling attractor solutions at late-time which can alleviate the coincidence problem.

\appendix
\section{The explicit form of the eignevalues of point $G\pm$} \label{B}

\begin{eqnarray}
g_1=\frac{3}{2}\frac{16\alpha^5-8\alpha^4\lambda-72\alpha^3+21\alpha^2\lambda+72\alpha-9\lambda}
{\alpha(8\alpha^4-21\alpha^2+9)}\hspace{0.5cm},\nonumber\\
\end{eqnarray}

\begin{widetext}
\begin{eqnarray}
&&g_2=\frac{3}{\Big(8\alpha^6-37\alpha^4+51\alpha^2-18\Big)\Big(8\alpha^4-21\alpha^2+9\Big)}
\bigg(-140\alpha^8+948\alpha^6-2322\alpha^4+2376\alpha^2-810+\nonumber\\
&&\Big(1024\alpha^{20}-30336\alpha^18+335616\alpha^{16}-1982094\alpha^{14}+7121691\alpha^{12}-16426287\alpha^{10}+24623838\alpha^{8}\nonumber\\
&&-23495670\alpha^{6}13434741\alpha^{4}-4056885\alpha^{2}+485514\Big)^{\frac{1}{2}}\bigg),\hspace{0.5cm}
\end{eqnarray}
\end{widetext}

\begin{widetext}
\begin{eqnarray}
&&g_3=\frac{3}{\Big(8\alpha^6-37\alpha^4+51\alpha^2-18\Big)\Big(8\alpha^4-21\alpha^2+9\Big)}
\bigg(140\alpha^8+948\alpha^6-2322\alpha^4+2376\alpha^2-810+\nonumber\\
&&-\Big(1024\alpha^{20}-30336\alpha^18+335616\alpha^{16}-1982094\alpha^{14}+7121691\alpha^{12}-16426287\alpha^{10}+24623838\alpha^{8}\nonumber\\
&&-23495670\alpha^{6}13434741\alpha^{4}-4056885\alpha^{2}+485514\Big)^{\frac{1}{2}}\bigg),\hspace{0.5cm}
\end{eqnarray}
\end{widetext}

\end{document}